\documentstyle[11pt,amssymb]{article}

\makeatletter
\def\citen#1{%
\edef\@tempa{\@ignspaftercomma,#1, \@end, }
\edef\@tempa{\expandafter\@ignendcommas\@tempa\@end}%
\if@filesw \immediate \write \@auxout {\string \citation {\@tempa}}\fi
\@tempcntb\m@ne \let\@h@ld\relax \def\@citea{}%
\@for \@citeb:=\@tempa\do {\@cmpresscites}%
\@h@ld}
\def\@ignspaftercomma#1, {\ifx\@end#1\@empty\else
   #1,\expandafter\@ignspaftercomma\fi}
\def\@ignendcommas,#1,\@end{#1}
\def\@cmpresscites{%
 \expandafter\let \expandafter\@B@citeB \csname b@\@citeb \endcsname
 \ifx\@B@citeB\relax 
    \@h@ld\@citea\@tempcntb\m@ne{\bf ?}%
    \@warning {Citation `\@citeb ' on page \thepage \space undefined}%
 \else
    \@tempcnta\@tempcntb \advance\@tempcnta\@ne
    \setbox\z@\hbox\bgroup 
    \ifnum0<0\@B@citeB \relax
       \egroup \@tempcntb\@B@citeB \relax
       \else \egroup \@tempcntb\m@ne \fi
    \ifnum\@tempcnta=\@tempcntb 
       \ifx\@h@ld\relax 
          \edef \@h@ld{\@citea\@B@citeB }%
       \else 
          \edef\@h@ld{\hbox{--}\penalty\@highpenalty
            \@B@citeB }%
       \fi
    \else   
       \@h@ld\@citea\@B@citeB
       \let\@h@ld\relax
 \fi\fi%
 \def\@citea{,\penalty\@highpenalty\hskip.13em plus.1em minus.1em}%
}
\def\@citex[#1]#2{\@cite{\citen{#2}}{#1}}%
\def\@cite#1#2{\leavevmode\unskip
  \ifnum\lastpenalty=\z@\penalty\@highpenalty\fi
  \ [{\multiply\@highpenalty 3 #1
      \if@tempswa,\penalty\@highpenalty\ #2\fi 
    }]\spacefactor\@m}
\makeatother

\let\a=\alpha \let\b=\beta

    \let\F=\Phi

\def\nn{\nonumber} \def\bd{\begin{document}} \def\ed{\end{document}}
\def\ds{\documentstyle} \let\fr=\frac \let\bl=\bigl \let\br=\bigr
\let\Br=\Bigr \let\Bl=\Bigl 
\let\bm=\bibitem
\let\na=\nabla
\let\pa=\partial \let\ov=\overline 
\newcommand{\be}{\begin{equation}} 
\newcommand{\ee}{\end{equation}} 
\def\ba{\begin{array}}
\def\ea{\end{array}}
\def\ft#1#2{{\textstyle{{\scriptstyle #1}\over {\scriptstyle #2}}}}
\def\fft#1#2{{#1 \over #2}}
\def\ffft#1#2{{ {#1}^{\phantom{A}}_{\phantom{\eta}} \over 
                 {#2}^{\phantom{A}}_{\phantom{\eta}} }}
\def\del{\partial}
\def\vp{\varphi}
\def\sst#1{{\scriptscriptstyle #1}}
\def\oneone{\rlap 1\mkern4mu{\rm l}}
\def\simequiv{\buildrel\sim\over=}
\def\td{\tilde}
\def\wtd{\widetilde}
\def\ie{{\it i.e.\ }}
\def\im{{\rm i}}
\def\dalemb#1#2{{\vbox{\hrule height .#2pt
        \hbox{\vrule width.#2pt height#1pt \kern#1pt
                \vrule width.#2pt}
        \hrule height.#2pt}}}
\def\square{\mathord{\dalemb{6.8}{7}\hbox{\hskip1pt}}}
\def\R{\rlap{\rm I}\mkern3mu{\rm R}}
\def\sR{\rlap{\hbox{$\scriptstyle\rm I$}}\mkern3mu{\hbox{$
\scriptstyle\rm R$}}}
\def\E{\rlap{\rm I}\mkern3mu{\rm E}}
\def\Z{\rlap{\sf Z}\mkern3mu{\sf Z}}
\def\F#1#2{{ F_{#1}^{(#2)} }}
\def\cF#1#2{{ {\cal F}_{#1}^{(#2)} }}
\def\ket#1{{|#1\rangle}}
\def\semi{\ltimes}
\newcommand{\ho}[1]{$\, ^{#1}$}
\newcommand{\hoch}[1]{$\, ^{#1}$}
\newcommand{\bea}{\begin{eqnarray}} 
\newcommand{\eea}{\end{eqnarray}} 
\newcommand{\ra}{\rightarrow}
\newcommand{\lra}{\longrightarrow}
\newcommand{\Lra}{\Leftrightarrow}
\newcommand{\ap}{\alpha^\prime}
\newcommand{\bp}{\tilde \beta^\prime}
\newcommand{\tr}{{\rm tr} }
\newcommand{\Tr}{{\rm Tr} } 
\newcommand{\NP}{Nucl. Phys. }

\newcommand{\sissatamphys}{\it SISSA, Via Beirut No. 2-4, 34013 Trieste, 
Italy and\\
Center for Theoretical Physics,
Texas A\&M University, College Station, Texas 77843}
\newcommand{\ens}{\it Laboratoire de Physique Th\'eorique de l'\'Ecole
Normale Sup\'erieure\hoch{3}\\
24 Rue Lhomond - 75231 Paris CEDEX 05}
\newcommand{\ic}{\it The Blackett Laboratory, Imperial College,\\
Prince Consort Road, London SW7 2BZ, UK}

\newcommand{\auth}{H. L\"u\hoch{\dagger}, C.N. Pope\hoch{\ddagger2}
and K.S. Stelle\hoch{\star}}

\begin{document}
\begin{flushright}
\hfill{CTP TAMU-35/97}\\
\hfill{Imperial/TP/96-97/59}\\
\hfill{LPTENS-97/38}\\
\hfill{SISSARef. 103/97/EP}\\
\hfill{hep-th/9708109}\\
\hfill{August 1997}\\
\end{flushright}

\begin{center}
{\Large{\bf Multiplet Structures of BPS Solitons}}\hoch{1}

\vspace{15pt}
\auth

\vspace{10pt}

{\hoch{\dagger}\ens}

\vspace{10pt}
{\hoch{\ddagger}\sissatamphys}

\vspace{10pt}
{\hoch{\star}\ic}

\vspace{40pt}

\underline{ABSTRACT}
\end{center}

   There exist simple single-charge and multi-charge BPS $p$-brane
solutions in the $D$-dimensional maximal supergravities.  From these,
one can fill out orbits in the charge vector space by acting with the
global symmetry groups.  We give a classification of these orbits, and
the associated cosets that parameterise them.

{\vfill\leftline{}\vfill
\vskip	10pt
\footnoterule
{\footnotesize \hoch{1} Research supported in part by the
European Commission under TMR contract \vskip -12pt} \vskip 10pt
{\footnotesize \hoch{\phantom{1}}
     ERBFMRX-CT96-0045. \vskip -12pt} \vskip 14pt
{\footnotesize	\hoch{2} Research supported in part by DOE 
Grant DE-FG03-95ER40917 and the \vskip -12pt} \vskip 10pt
{\footnotesize \hoch{\phantom{2}} EC Human Capital and Mobility 
Programme under contract ERBCHBGT920176.\vskip	-12pt} \vskip 14pt
{\footnotesize
        \hoch{3} Unit\'e Propre du Centre National de la Recherche
Scientifique, associ\'ee \`a l'\'Ecole Normale Sup\'erieure.
\vskip -12pt} \vskip 10pt
{\footnotesize \hoch{\phantom{3}} et \`a l'Universit\'e de Paris-Sud
\vskip -12pt} \vskip 10pt}

\pagebreak
\setcounter{page}{1}

\section{Introduction}

     The BPS $p$-brane solutions in supergravity theories play a
central r\^ole in the understanding of the non-perturbative structures
of M-theory and string theory.  For this reason, it is useful to
develop a framework for characterising these solutions, and in
particular, for describing their multiplet structures under the
duality symmetries of the theories.  In this paper, we shall be
principally concerned with studying the multiplet structures of BPS
solitons in maximal supergravities.  In particular, we shall study the
orbits of the global supergravity symmetry groups $E_{11-D}$ in $D$
dimensions \cite{cj1,cj2},\footnote{In all cases, unless otherwise
indicated, it is to be understood that the groups that we encounter
will be of the maximally non-compact form, $E_{n(n)}$, {\it etc}.  We
shall in general omit the explicit indication of the maximal
non-compact form.}  to see how they fill out multiplets of solutions.

    A systematic way to study the orbits is first to consider the
simplest single-charge BPS $p$-brane solutions, which preserve $\ft12$
of the supersymmetry.  Such a solution is described by a single
harmonic function in the space transverse to the world-volume of the
$p$-brane.  Acting with the global symmetry group $G$, one obtains
more complicated solutions in which, typically, more than one charge
is non-vanishing.  Of course since the global symmetry transformations
$G$ commute with supersymmetry, all the solutions in the orbit will
also preserve $\ft12$ of the supersymmetry.  All the solutions in the
orbit are characterised by the same single harmonic function.  In some
cases these orbits fill out the entire charge vector-space. Under
these circumstances, all points in the charge vector space are
associated with BPS solutions that preserve $\ft12$ the supersymmetry.

    In other cases, it turns out that the orbits of the original
single-charge solution fill out only a subspace in the entire charge
vector space.  The signal for this is that there can be points in the
charge space, in the neighbourhood of the single-charge starting
point, which do not lie on the orbit.  When this occurs, it turns out
that there exists a ``simple'' 2-charge solution, which has both the
original charge and any one of the charges that lies outside the
infinitesimal orbit of the original charge.  Here, by a simple
2-charge solution, we mean one that is characterised by two
independent harmonic functions, associated with exactly two
non-vanishing charges.\footnote{In this paper, we shall use the term
``simple $N$-charge solution'' to mean a solution with $N$ independent
harmonic functions that are associated with exactly $N$ non-vanishing
charges.  Such solutions form multiplets under the Weyl group of the
global symmetry group $G$\cite{lpsweyl}; they were classified in
\cite{classp}.  Acting with $G$ on a simple $N$-charge solution, one
always obtains configurations with $N'\ge N$ non-vanishing charges
\cite{classp}.  Equality is achieved only for Weyl-group elements.}
All such 2-charge solutions preserve $\ft14$ of the supersymmetry.
One can now repeat the exercise of filling out the orbits under $G$,
taking the simple 2-charge solution as a new starting point.  Again,
there is the possibility that the orbits now cover the entire charge
vector space, in which case the multiplet analysis for this class of
solution is complete.  Alternatively, it may be the case that there
are still points in the neighbourhood of the simple 2-charge solution
that cannot be reached by the orbit.  This indicates the existence of
a simple 3-charge solution.  By iteratively applying this procedure of
adding in further ``missed'' charges, one eventually finds orbits that
do cover the entire charge space.  It turns out that for solutions
whose charges are carried by the 4-form field strength, simple
1-charge solutions are always sufficient to cover the entire charge
vector space.  For solutions supported by 3-forms and 2-forms, the
maximal values of $N$ for simple $N$-charge solutions are dimension
dependent; they were all obtained in \cite{lpsol,lpmult}, and
classified in \cite{classp}.  For 3-forms, one always has $N\le2$, and
for 2-forms, $N\le 4$.  The different orbits can also be characterised
by certain group-invariant polynomials, as discussed in \cite{fm}.

    It should be emphasised that the problem that we are addressing in
this paper, of studying the action of the global symmetry groups on
the charge vectors in BPS solutions, is not the same as the problem of
studying the spectrum-generating symmetries for BPS solutions.  To
give the spectrum of BPS solitons, one needs to classify the sets of
solutions at {\it fixed} values of the scalar moduli, \ie the
asymptotic values of all the dilatonic and axionic scalars.  Although
the standard global symmetry algebras are effective in mapping between
different points in the charge vector space, they also in general
change the scalar moduli at the same time.  This problem was addressed
in \cite{clps}, in the case of the spectrum obtained from
single-charge BPS solutions that preserve $\ft12$ of the
supersymmetry.  Since the spectrum for fixed moduli necessarily
involves solutions with different masses, it is manifest that the
standard global symmetry groups cannot generate the complete spectrum.
As was shown in \cite{clps}, the additional ingredient of the overall
scaling symmetry that every supergravity theory possesses is needed
also, in order to fill out the entire spectrum.  (The scaling
transformation, called the ``trombone'' transformation in \cite{clps},
is a symmetry of the equations of motion, corresponding to an
homogeneous scaling of the action.)  It turns out that the spectrum
generating symmetry for the solutions that preserve $\ft12$ the
supersymmetry is in fact the same group as the global symmetry group
$G$, but now realised non-linearly on the fields.  We shall return to
a discussion of this point in section 9, and in particular address the
question of whether one can expect the results in \cite{clps} to
extend to the multi-charge solutions that preserve less than $\ft12$
of the supersymmetry.

    In this paper, we shall study the orbits of the charge vectors for
all the $p$-branes in $D\ge4$ that are supported by field strengths of
degree 4, 3 and 2.  All of these field strengths form linear
representations under the global symmetry groups.  In all cases, the
dimensions of the global symmetry groups are larger than the
dimensions of the charge vector spaces.  In other words, there is a
stability subgroup $K$ of $G$ that leaves the initial simple
$N$-charge configuration fixed.  The orbits are therefore
parameterised by points in the coset $G/K$.  We shall determine these
coset structures for all the above $p$-brane orbits.  (We shall focus
on the classical coset structure in this paper.  At the quantum level,
the continuous global symmetry groups will be discretised to the
U-duality groups discussed in \cite{ht}.)  The coset for the type IIB
string doublet was obtained in \cite{clps}, and the cosets for $D=5$
and $D=4$ have recently been obtained in \cite{fg}.

     In order to describe the maximal supergravities in $D$
dimensions, we shall use the notation and conventions of \cite{lpsol}.
The $D$-dimensional  Lagrangian is given by
\bea
{\cal L} &=& eR -\ft12 e\, (\del\vec\phi)^2 -\ft1{48}e\, e^{\vec a\cdot
\vec\phi}\, F_4^2 -\ft{1}{12} e\sum_i
e^{\vec a_i\cdot \vec\phi}\, (F_3^{(i)})^2
-\ft14 e\, \sum_{i<j} e^{\vec a_{ij}\cdot \vec\phi}\, (F_2^{(ij)})^2
\label{dgenlag}\\
&& -\ft14e\, \sum_i e^{\vec b_i\cdot \vec\phi}\, ({\cal F}_2^{(i)})^2
-\ft12 e\, \sum_{i<j<k} e^{\vec a_{ijk} \cdot\vec \phi}\,
(F_1^{(ijk)})^2 -\ft12e\, \sum_{i<j} e^{\vec b_{ij}\cdot \vec\phi}\,
({\cal F}_1^{(ij)})^2 + {\cal L}_{\sst{FFA}}\ ,\nn
\eea
where the subscripts on the various field strengths indicate their
degrees, with $F_4$, $\F3i$, $\F2{ij}$ and $\F1{ijk}$ coming from the
4-form in $D=11$, while $\cF2i$ and $\cF1{ij}$ come from the
dimensional reduction of the metric.  The ``dilaton vectors'' $\vec
a$, $\vec a_i$, $\vec a_{ij}$, $\vec a_{ijk}$, $\vec b_i$ and $\vec
b_{ij}$ are constant $(11-D)$-component vectors that characterise the
couplings of the dilatonic scalars $\vec \phi$ to the various field
strengths.  Their detailed forms, and also their scalar products, are
all given in \cite{lpsol}:
\bea
&&F_{\sst{MNPQ}}\qquad\qquad\qquad\qquad\qquad\qquad\qquad\qquad
{\rm vielbein}\nonumber\\
{\rm 4-form:}&&\vec a = -\vec g\ ,\nonumber\\
{\rm 3-forms:}&&\vec a_i = \vec f_i -\vec g \ ,\nonumber\\
{\rm 2-forms:}&& \vec a_{ij} = \vec f_i + \vec f_j - \vec g\ ,
\qquad\qquad\qquad\qquad\qquad \,\,\, \,\vec b_i = -\vec f_i \ ,
\label{dilatonvec}\\
{\rm 1-forms:}&&\vec a_{ijk} = \vec f_i + \vec f_j + \vec f_k -\vec g
\ ,\qquad\qquad\qquad\qquad\vec b_{ij} = -\vec f_i + \vec f_j\ .\nonumber 
\nonumber
\eea
The explicit expressions for the vectors $\vec g$ and $\vec f_i$, which 
have $(11-D)$ components in $D$ dimensions, are given in \cite{lpsol}.  For
our purposes, it is sufficient to note that they satisfy the relations
\be
\vec g \cdot \vec g = \ft{2(11-D)}{D-2}, \qquad
\vec g \cdot \vec f_i = \ft{6}{D-2}\ ,\qquad
\vec f_i \cdot \vec f_j = 2\delta_{ij} + \ft2{D-2}\ .\label{gfdot}
\ee

    The scalar sector of the $D$-dimensional theory comprises the
$(11-D)$ dilatons $\vec \phi$, and the axions $A_0^{(ijk)}$ and ${\cal
A}_0^{(ij)}$ whose field strengths are the 1-forms $\F1{ijk}$ and
$\cF1{ij}$.  In addition, in the conventionally ``fully-dualised''
versions of the theories, where all field strengths of degree $>\ft12
D$ are dualised, there will be additional axions in $D\le5$ associated
with the dualisation of $(D-1)$-form field strengths.  Their
corresponding dilaton vectors will be the negatives of those for the
fields prior to dualisation.  Including the axions coming from
dualisation, the total number of scalars is equal to the dimension of
the Borel subgroup of the $E_{11-D}$ Cremmer-Julia global symmetry
group \cite{cjlp}.  In fact, the dilatons $\vec\phi$ are in one-to-one
correspondence with the Cartan subalgebra, while the axions are in
one-to-one correspondence with the positive roots of $E_{11-D}$.  In
particular, their dilaton vectors are precisely the positive root
vectors.  The simple roots in $D$ dimensions may be taken to be $\vec
b_{i,i+1}$ and $\vec a_{123}$, for $1\le i \le 10-D$.  It is easy to
verify from (\ref{dilatonvec}) and (\ref{gfdot}) that these vectors
have the dot products of the simple roots of $E_{11-D}$, as shown in
Table 1.

\bigskip\bigskip

\centerline{
\begin{tabular}{ccccccccccccc}\\
 $\vec b_{12}$& &$\vec b_{23}$& &$\vec b_{34}$& &$\vec b_{45}$
& &$\vec b_{56}$& &$\vec b_{67}$& &$\vec b_{78}$ \\
 o&---&o&---&o&---&o&---&o&---&o&---&o\\
 &   & &   &$|$&   & &   & &   & &   & \\
 &   & &   &o&   & &   & &   & &   & \\
 &   & &   &$\vec a_{123}$& & &   & &   & &   & \\
\end{tabular}}
\bigskip\bigskip

\centerline{Table 1: The dilaton vectors $\vec b_{i,i+1}$ and $\vec
a_{123}$ generate the $E_{11-D}$ Dynkin diagram}
\bigskip\bigskip
In each dimension $D$, the diagram is truncated to the part that survives
when only the simple roots with indices $i\le 11-D$ are retained.

    In the fully-dualised theories, the higher-degree fields form
linear highest-weight representations under the $E_{11-D}$ group.  The
associated dilaton vectors are the weight vectors of the
representation \cite{lpsweyl}.  The action of the generators of the
global symmetry group on the states in the various representations can
easily be determined by means of the dilaton vectors.  The dilaton
vectors for the axions are the positive roots of the $E_{11-D}$
algebra (in the fully-dualised theory).  The group generator
associated with a root $\vec \a$ gives a non-vanishing result when
acting on a state of weight $\vec w$ if and only if $\vec w+\vec\a$ is
also a weight in the representation.  We shall make repeated use of
this in what follows.  All of the necessary results can be derived
from the information about the dilaton vectors that is contained in
(\ref{dilatonvec}) and (\ref{gfdot}).

     The paper is organised as follows.  Sections 2 to 7 contain the
discussion of the charge vector orbits for field strengths with
degrees $\ge 2$ in all dimensions between $D=10$ and $D=4$.  In each
case, we analyse all the multi-charge starting points that are needed
in order to fill out the orbits that span the entire charge spaces,
and we discuss the supersymmetry of the various classes of solution.
Our results in $D=5$ and $D=4$ are similar to those obtained in
\cite{fg}.  In section 8, we consider the extension of these ideas to
solutions where the charges are carried by 1-form field strengths.
Owing to the non-linearity of the action of the global symmetry groups
on these fields, a general discussion is much more complicated than in
the higher-degree cases.  Instead, we consider a simplified situation
where a subset of the axions form a linear representation under a
subgroup of the full global symmetry group.  The paper ends with
concluding remarks in section 9.  A summary of the results for the
various cosets is contained in an appendix.
   
\section{$D=10$ Type IIB and $D=9$}

     In both of these cases, there is an $SL(2,\R)$ global
symmetry. Let us first consider the type IIB theory.  This contains
two 3-form field strengths, which transform as a doublet under
$SL(2,\R)$, and a singlet self-dual 5-form, together with the dilaton
and axion that parameterise the $SL(2,\R)/SO(2)$ scalar manifold.  The
BPS string solutions, with electric charges $(Q_1,Q_2)$ for the two
3-form field strengths, form a single multiplet under $SL(2,\R)$.  For
example, one can start from the solution with charges $(1,0)$, and
fill out the entire 2-dimensional charge space by acting with
$SL(2,\R)$.

     In fact, we should pause at this point to make this statement
more precise.  At the same time as the standard $SL(2,\R)$ symmetry
transforms the charges, it also changes the scalar moduli, \ie the
asymptotic values of the dilaton and axion.  What we are really
interested in is to fill out the 2-dimensional charge space at {\it
fixed} values of the moduli.  In particular, this involves BPS
solutions with different masses, and it is manifest that the standard
$SL(2,\R)$ symmetry cannot by itself achieve this, since it leaves the
metric invariant and hence preserves the mass.  As was recently
discussed in \cite{clps}, the true spectrum-generating symmetry that
fills out the 2-dimensional charge space at fixed values of the scalar
moduli is obtained by including also the scaling symmetry of the
theory, which of course {\it does} change the mass.  To be precise,
the charge space can generated by acting with the denominator subgroup
(which is $SO(2)$ in this case) to rotate the original charge vector
$(1,0)$ keeping its length fixed, and then acting with the rescaling
``trombone'' symmetry to arrive at the desired radius in the
2-dimensional charge plane.  An alternative way to describe this,
which allows a proper treatment of the effects of Dirac quantisation
in the quantum case, is that we act with the standard $SL(2,\R)$ on
the charge space, and then apply compensating symmetry transformations
that leave the charges fixed, but restore the scalar moduli to their
former values.  Specifically, this is done by acting with a
combination of a transformation in the Borel subgroup of $SL(2,\R)$,
together with a trombone rescaling.  The Borel
transformation\footnote{The Iwasawa decomposition of the $E_{11-D}$
global symmetry groups into the product of the Borel and maximal
compact subgroups has been studied in \cite{decomp}.}  allows us to
restore the scalar moduli to their original values, but at the price
that the charges are rescaled (although their ratio remains fixed).
The trombone transformation then restores the scale of the charges,
while leaving the moduli at their proper restored values.  The net
effect of this is that we now have a {\it non-linear} realisation of
$SL(2,\R)$, which acts as a true spectrum generating symmetry
\cite{clps}.  In the quantum case, this spectrum-generating $SL(2,\R)$
is discretised to $SL(2,\Z)$, in order to ensure that the Dirac
quantisation condition is satisfied.

     The results in \cite{clps} established that the group of the true
spectrum-generating symmetry is the same, {\it qua} abstract group, as
the standard Cremmer-Julia symmetry group, but now realised
non-linearly on the fields.  However, its realisation on the charges
is still linear, and so for our present purposes we may continue to
think of the standard linearly-realised Cremmer-Julia group as the
group that fills out multiplets in the vector space of the charges,
with the understanding that the full description of the associated BPS
solutions would require compensating transformations on the scalar
moduli.  We shall return to a discussion of this point in section 9.

     The group-theoretic structure of the spectrum of the type IIB
BPS string solitons was discussed in \cite{clps}.  The $SL(2,\R)$
transformations 
\be
\pmatrix{Q_1\cr Q_2} = \pmatrix{a&b\cr c&d} \pmatrix{1\cr0} =
\pmatrix{a\cr c}\label{sl2r}
\ee
do not all act effectively on the initial charge vector $(1,0)$; in
fact the strict Borel subgroup
\be
B_{\rm strict} = \pmatrix{1&k\cr 0&1}\label{strict}
\ee
leaves the initial charge vector $(1,0)$ fixed.  Therefore the points
in the 2-dimensional charge space are parameterised by elements of the
coset $SL(2,\R)/B_{\rm strict}$.  Clearly matrices of the form
(\ref{strict}) give a realisation of $\R$, and so we may equivalently
say that the charge space for the
BPS strings in the type IIB theory is parameterised by \cite{clps}
\be 
\fft{SL(2,\R)}{\R}\ .
\ee
The initial single-charge solution, and the solutions corresponding to
all other points in the charge space, preserve $\ft12$ of the
supersymmetry.

     An analogous discussion can be given for the multiplet of
5-branes, corresponding to the case where the two 3-form field
strengths carry magnetic charges.  These charges also transform as a
doublet under $SL(2,\R)$, contragrediently with respect to the
transformations of the electric charges.  Thus again, the charge space
is parameterised by the coset $SL(2,\R)/\R$, and again all points are
associated with BPS solutions that preserve $\ft12$ of the
supersymmetry.
 
     There also exist self-dual 3-branes in the type IIB theory,
supported by the self-dual 5-form.  This is a singlet under
$SL(2,\R)$.  However, we know that BPS 3-brane solitons can exist for 
arbitrary values of the charge (classically speaking).  These
solutions can all be obtained from a solution with a given charge, by
acting with the trombone scaling transformation.  (Such transformations
are also responsible for the M-brane solutions in $D=11$ occurring for
arbitrary values of the charge.) 

    Turning now to maximal supergravity in nine dimensions, we now
have a $GL(2,\R)=SL(2,\R)\times \R$ global symmetry (together also
with a trombone scaling symmetry).  There is an $SL(2,\R)$ doublet of
3-form field strengths $\F3i$, and an $SL(2,\R)$ doublet of 2-form
field strengths $\cF2i$.  In addition, there is a singlet 4-form
$F_4$, a singlet 2-form $\F2{12}$, two dilatonic scalars and an
axionic scalar.  Here, we shall look at the group structure of the
various BPS solitons under the $SL(2,\R)$ group.
 
     Since the 4-form is a singlet under $SL(2,\R)$, the only way to
get an arbitrary-charge solution from a given one is by means of the
trombone symmetry.  The analysis of the spectrum of BPS strings or
4-branes for a doublet of 3-forms $\F3i$ is identical to that in the
type IIB case, and again the 2-dimensional charge space is
parameterised by the coset $SL(2,\R)/\R$.  

     The discussion for the 2-form solutions is slightly more
complicated.  There are three 2-forms in total, namely the $SL(2,\R)$
doublet $\cF2i$ and the singlet $\F2{12}$.  If we start from a
single-charge solution using one member of the doublet $\cF2i$, then
acting with $SL(2,\R)$ can only fill out the 2-dimensional plane
$(p_1,p_2,0)$ in the the 3-dimensional charge space
$(p_1,p_2,u_{12})$.  However, there also exist 2-charge solutions
using the field strengths $\{\cF21,\F2{12}\}$ or $\{\cF22,\F2{12}\}$.
If, for example, we start from the charge configuration
$(1,0,u_{12})$, where $u_{12}$ is any fixed charge for $\F2{12}$, then
acting with $SL(2,\R)$ will fill out the entire 2-plane
$(p_1,p_2,u_{12})$ at the fixed value of $u_{12}$.  Since $u_{12}$ can
be given an arbitrary value, it follows that any point in the
3-dimensional charge space is associated with a $p$-brane solution.
Those with charges $(p_1,p_2,0)$ or $(0,0,u_{12})$ preserve $\ft12$
the supersymmetry, whilst those with charges $(p_1,p_2,u_{12})$ with
at least one of $p_1$ or $p_2$ non-zero preserve $\ft14$.

     We have seen in the above discussion that in certain cases,
namely the 3-form solutions in type IIB or in $D=9$, BPS solutions
corresponding to any point in the 2-dimensional charge space are
filled out by acting with the global symmetry group on a basic
single-charge solution, which preserves $\ft12$ of the supersymmetry.
On the other hand, the full set of BPS 2-form solutions in $D=9$ is
filled out by starting with a basic 2-charge solution, and acting with
the global symmetry group.  Note that this 2-charge solution preserves
$\ft14$ of the supersymmetry, and the number of charges cannot be
reduced to one by any global symmetry transformation.  In fact, global
symmetry transformations only yield solutions with $N\ge2$ charges.
For this reason, it is useful to define the basic 2-charge solution as
being a {\it simple} 2-charge solution.  The full classification of
all simple $N$-charge $p$-branes in all dimensions $D\ge2$ was given
in \cite{classp}.
  
\section{$D=8$}

    The global symmetry group in $D=8$ is $SL(3,\R)\times SL(2,\R)$,
where the first factor has $\vec b_{12}$ and $\vec b_{23}$ as simple
roots, and the second has the simple root $\vec a_{123}$ (see Table
1).  The full root system is given by $\pm\vec b_{ij}$ and $\pm \vec
a_{123}$.  There is one 4-form field strength $F_4$, which, together
with its dual, forms a doublet under the $SL(2,\R)$.  The three
3-forms $\F3i$ form a triplet under $SL(3,\R)$, and the six 2-forms
$\F2{ij}$ and $\cF2i$ form a (3,2) irreducible representation under
$SL(3,\R)\times SL(2,\R)$.  The associated dilaton vectors are the
weight vectors of these various irreducible representations.  They,
and the highest-weight vectors, are presented in Table 2.  Also
included is the list of negative roots whose generators annihilate the
highest-weight states (of course all the positive roots, by
definition, annihilate the highest weights).

\bigskip\bigskip\bigskip\bigskip

\centerline{
\begin{tabular}{|c|c|c|c|}\hline
Degree & Weight vectors & Highest weight &
$\begin{array}{c}\hbox{Negative roots that} \\ 
\hbox{annihilate highest weight}
\end{array}$ 
\\ \hline\hline
4-form & $\vec a, -\vec a$  & $ -\vec a$ & $ -\vec b_{ij}$\\ \hline
3-forms& $\vec a_i$         & $\vec a_3$ & $-\vec b_{12}, -\vec
a_{123}$ \\ \hline
2-forms& $\vec a_{ij}, \vec b_i$ & $\vec a_{23}$ & $-\vec b_{23}$\\ \hline   
\end{tabular}}
\bigskip\bigskip

\centerline{Table 2: Weights and highest weights in $D=8$}
\bigskip\bigskip

     Since the electric and magnetic charges of the 4-form form a
doublet under $SL(2,\R)$, the discussion of its multiplet structure
is identical to that for the $SL(2,\R)$ multiplets in the previous
section.  The points in the 2-dimensional electric/magnetic charge
space are parameterised by $SL(2,\R)/\R$.  The $\R$ denominator
group is the strict Borel subroup of the $SL(2,\R)$, whose generator
has the positive root $\vec a_{123}$.  

     The highest-weight state $\ket{\vec a_3}$ representing the
single-charge 3-form solution is annihilated not only by the
positive-root generators $E_{\vec b_{ij}}$ and $E_{\vec a_{123}}$, but
also by $E_{-\vec b_{12}}$ and $E_{-\vec a_{123}}$.  This means that
only the negative-root generators $E_{-\vec b_{13}}$ and $E_{-\vec
b_{23}}$ have a non-vanishing action on the highest-weight state,
giving $\ket{\vec a_1}$ and $\ket{\vec a_2}$ respectively.  In
addition, there is one combination of Cartan generators under which
the state $\ket{\vec a_3}$ has non-vanishing weight (the combination
proportional to $\vec a_3\cdot \vec H$).  Thus there is a
three-parameter family of motions of the highest-weight state.  In
other words, the coset space that parameterises the 3-form solutions
obtained from the global group action on the single charge associated
with $\ket{\vec a_3}$ is three dimensional.  The stability subgroup is
generated by $E_{\vec b_{ij}}$ and $E_{\vec a_{123}}$, together with
$E_{-\vec b_{12}}$ and $E_{-\vec a_{123}}$ and the two remaining
combinations of the Cartan generators,
\be
\vec H' = \vec H -\ft34 (\vec a_3\cdot\vec H)\, \vec a_3\ .\label{d8h}
\ee
The
vectors $\pm \vec a_{123}$ and $\pm \vec b_{12}$ are the non-vanishing
roots of two commuting $SL(2,\R)$ groups, generated by $E_{\pm\vec a_{123}}$
and  $E_{\pm\vec b_{12}}$, whose two Cartan generators
are the projections of $\vec H'$ onto the mutually-orthogonal
directions $\vec a_{123}$ and $\vec b_{12}$ respectively (the roots of
the two commuting $SL(2,\R)$ groups 
are unchanged by the projection (\ref{d8h}) onto $\vec H'$, since
$\vec a_3\cdot \vec a_{123}=\vec a_3\cdot\vec b_{12}=0$). The
remaining generators, $E_{\vec b_{13}}$ and $E_{\vec b_{23}}$, are
mutually commuting, and they commute with
$E_{\pm\vec a_{123}}$, but form a doublet under the $SL(2,\R)$
generated by $E_{\pm\vec b_{12}}$, with weight vectors 
\be
\pm\ft12(\vec b_{13}+\vec b_{23})
\ee 
under $\vec H'$.  Thus the structure of the
3-dimensional coset space parameterising the 3-charge solutions is
\be
\fft{SL(3,\R)\times SL(2,\R)}{(SL(2,\R)\semi \R^2)\times SL(2,\R)}\ .
\ee
Since the $SL(2,\R)$ factor generated by $E_{\pm\vec a_{123}}$ is common
to the numerator and denominator, the coset reduces to
\be
\fft{SL(3,\R)}{SL(2,\R)\semi \R^2}\ ,\label{d83f}
\ee
where the $SL(2,\R)$ is generated by $E_{\pm\vec b_{12}}$, with roots
$\pm\vec b_{12}$, and the two commuting $\R$ generators have weight
vectors $\vec b_{13}$ and $\vec b_{23}$.  The fact that they transform
as a doublet under $SL(2,\R)$ is reflected in the semi-direct product
symbol $\semi$.

     Note that the dimension of the coset (\ref{d83f}) is equal to the
dimension of the space of 3-form charges, \ie 3.  This is not a
coincidence; the 3-forms transform as the vector representation of
$SL(3,\R)$, and so any point in the charge space can be reached by
acting with $SL(3,\R)$ on a given charge, which we are taking to be
associated with the highest-weight state $\ket{\vec a_3}$. Thus any
point in the 3-dimensional charge space has an associated BPS string
or 3-brane solution, which preserves $\ft12$ of the supersymmetry.

   Finally in this section, we consider the 2-form field strengths,
which form a (3,2)-dimensional irreducible representation under
$SL(3,\R)\times SL(2,\R)$, with $\vec a_{23}$ as the highest weight.
First, we consider the multiplet filled out by acting on the
single-charge solution associated with this highest-weight state.
From Table 2, we see that $\ket{\vec a_{23}}$ is annihilated by
$E_{\pm \vec b_{23}}$, which form the non-zero-root generators of an
$SL(2,\R)$, and by $E_{\vec b_{12}}$, $E_{\vec b_{13}}$ and $E_{\vec
a_{123}}$.  These are mutually-commuting generators, with ($E_{\vec
b_{12}}, E_{\vec b_{13}}$) transforming as a doublet under the
$SL(2,\R)$, and $E_{\vec a_{123}}$ as a singlet.  In addition, there
are two combinations of the Cartan generators that leave $\vec a_{23}$
invariant, namely
\be
\vec H'=\vec H -\ft37(\vec a_{23}\cdot\vec H)\, \vec a_{23}\ .
\ee
One combination is the Cartan generator $\vec b_{23}\cdot\vec H'$ in the
$SL(2,\R)$ subgroup generated by $E_{\pm\vec b_{23}}$, and the other,
which commutes with every generator in the stability group, enlarges
the $SL(2,\R)$ to $GL(2,\R)$.  Thus the coset space that parameterises
the orbits of the single-charge 2-form solutions is
\be
\fft{SL(3,\R)\times SL(2,\R)}{GL(2,\R)\semi \R^3}\ .\label{d82f1}
\ee
In fact it is easy to understand the structure of the denominator
group; there is a $GL(2,\R)$ global symmetry in $D=9$, which clearly,
since it leaves the $D=9$ metric invariant, will also leave the new
Kaluza-Klein vector ${\cal A}_1^{(3)}$ in $D=8$ invariant.  In
addition, the three vector potentials $\{ {\cal A}_1^{(1)}, {\cal
A}_1^{(2)}, A_1^{(12)}\}$ in $D=9$ give rise to three axions in $D=8$,
which have three commuting global $\R$ symmetries (two of which form a
doublet under $GL(2,\R)$).  Thus the full invariance group for a
solution using the new Kaluza-Klein vector is $GL(2,\R)\semi \R^3$.
Since ${\cal A}_1^{(3)}$ is equivalent to the previous potential
$A_1^{(23)}$ associated with the highest-weight vector $\vec a_{23}$
under the full global symmetry group $SL(3,\R)\times SL(2,\R)$, it
follows that the solution based on the field strength associated with
$\ket{\vec a_{23}}$ will also have $GL(2,\R)\semi \R^3$ as its
stability group.

     The dimension of the coset space (\ref{d82f1}) is 4, while the
dimension of the charge space for 2-forms is 6.  This shows that not
all points in the total charge space can be reached by acting with
global symmetry transformations on a single-charge solution.  In
particular, if we start from the single charge associated with the
state $\ket{\vec a_{23}}$, we will never generate the charges
associated with the states $\ket{\vec b_2}$ and $\ket{\vec b_3}$. In
fact, there also exist simple 2-charge solutions for 2-forms in $D=8$,
and we can choose to augment the original charge associated with $\vec
a_{23}$ by precisely one or other of the charges that the
single-charge orbits have failed to cover.  For example, we may
consider the case where $\F2{23}$ and $\cF23$ carry the charges
\cite{lpsol}.  Thus to reveal the structure of the coset space
parameterising these solutions, we need to identify the simultaneous
stability subgroup for the pair of states $\ket{\vec a_{23}}$ and
$\ket{\vec b_3}$.  To do this, it is useful to tabulate the action of
the complete set of all the non-zero-root generators of
$SL(3,\R)\times SL(2,\R)$ on the two states.  Labelling the states by
their weights, we have:

\bigskip\bigskip

\centerline{
\begin{tabular}{|c|c|c|c|c|c|c|c|c|}\hline
Weight & $\vec b_{12}$ & $\vec b_{13}$& $\vec b_{23}$& $\vec a_{123}$
& $-\vec b_{12}$ & $-\vec b_{13}$& $-\vec b_{23}$& $-\vec a_{123}$
\\ \hline\hline
$\vec a_{23}$ & -- & -- & -- & -- & $\vec a_{13}$ & $\vec a_{12}$ & --
& $\vec b_1$ \\ \hline
$\vec b_3$ &-- & $\vec b_1$ & $\vec b_2$ & $\vec a_{12}$ & -- & -- & --
& -- \\ \hline
\end{tabular}}
\bigskip\bigskip

\centerline{Table 3: Action of generators on 
$\ket{\vec a_{23}}$ and $\ket{\vec b_3}$ in $D=8$}
\bigskip\bigskip

  We see from Table 3 that the generators $E_{\vec b{12}}$ and
$E_{-\vec b_{23}}$ leave both of the states $\ket{\vec a_{23}}$ and
$\ket{\vec b_3}$ invariant.  In addition, we see that the combinations
of generators
\be
L_+=(E_{\vec b_{13}} - E_{-\vec a_{123}})\ ,\quad 
L_-=(E_{-\vec b_{13}} - E_{\vec a_{123}})
\ee 
will leave the sum of states $\ket{\vec a_{23}} + \ket{\vec
b_3}$ invariant.  These form the positive and negative root
generators of an $SL(2,\R)$ algebra.  The Cartan generator for this
$SL(2,\R)$ must come from the one combination of the three Cartan
generators of the original $SL(3,\R)\times SL(2,\R)$ under which both
the states $\ket{\vec a_{23}}$ and $\ket{\vec b_3}$ have zero weight,
namely
\be
\vec H'=\vec H -\ft18(5\vec a_{23}\cdot H + 7\vec b_3\cdot \vec H)\,
\vec b_3 -\ft18(5\vec b_3\cdot\vec H + 7 \vec a_{23}\cdot\vec H)\,
\vec a_{23}\ .
\ee
(The specific coefficients that arise here in this Gramm-Schmidt
projection follow from the details of the dot products of the dilaton
vectors involved in the construction.  These dot products can all be
determined from the results given in (\ref{dilatonvec}) and
(\ref{gfdot}).)  Thus the new Cartan generator $H'$ is the projection
of the original 3-vector $\vec H$ of Cartan generators onto the line
orthogonal to the plane containing the weights $\vec a_{23}$ and $\vec
b_3$ associated with the two charges.  Note that although the
generators $L_\pm$ are not eigenstates under the original Cartan
generators $\vec H$, they are eigenstates under the projected
generators $\vec H'$.  After a calculation, one finds that the two
$SL(2,\R)$ generators $L_\pm$ have weights $\pm\ft12(\vec b_{13}-\vec
a_{123})$ under $H'$.  In total, we therefore have five generators
that leave the two-charge combination invariant.  Clearly $L_+, L_-$
and the Cartan combination $\vec H'$ form an $SL(2,\R)$ algebra, and
the mutually commuting pair $E_{\vec b_{12}}$ and $E_{-\vec b_{23}}$
form a doublet under this $SL(2,\R)$.  Thus the stability subgroup of
the 2-charge combination is $SL(2,\R)\semi \R^2$, and the space of
2-charge solutions is parameterised by the coset
\be
\fft{SL(3,\R)\times SL(2,\R)}{SL(2,\R)\semi \R^2}\ .\label{d82f2}
\ee
Note that this space has dimension 6, which is the same as the total 
dimension of the space of 2-form charges in $D=8$.  In fact, this
coset allows us to reach all points in the 6-dimensional charge space,
beginning from a given 2-charge solution. This can be seen from
the non-trivial entries in Table 3, which indicate that all six of the 
states corresponding to the six 2-forms can be reached from the
2-charge starting point.  (The Cartan combinations $\vec a_{23}\cdot
\vec H$ and $\vec b_3\cdot\vec H$ allow the original 2 charges to be
rescaled arbitrarily, and the various combinations of non-zero-root
generators that do not annihilate $\ket{\vec a_{23}}$ or $\ket{\vec
b_3}$ allow the other four charges to be turned on with arbitrary
strengths.)  In general, the BPS solutions will preserve $\ft14$ of the
supersymmetry, unless one or other of the original charges is zero, in
which case the solution will preserve $\ft12$ of the supersymmetry.

\section{$D=7$}

     The global symmetry group in $D=7$ is $SL(5,\R)$, with simple
roots $\vec b_{12}$, $\vec b_{23}$, $\vec b_{34}$ and $\vec a_{123}$.
The full root system is given by $\pm\vec b_{ij}$ and $\pm \vec
a_{ijk}$, associated with the generators $E_{\pm\vec b_{ij}}$ and
$E_{\pm\vec a_{ijk}}$.  In the fully-dualised theory there are five
3-forms $\F3i$ and $*F_4$, and ten 2-forms $\F2{ij}$ and $\cF2i$.  The
dilaton vectors associated with the 3-forms and the 2-forms are the
weight vectors of the 5 and the $\bar{10}$ of $SL(5,\R)$ respectively.

     Let us begin by considering the 3-forms.  Here, the
highest-weight vector is $-\vec a$, associated with the 3-form $*F_4$.
It is easy to see the state $\ket{-\vec a}$ is annihilated not only by
all the positive-root generators, but also by the negative-root
generators associated with the roots $-\vec b_{ij}$, and by the three
combinations of the $SL(5,\R)$ Cartan generators that are orthogonal
to $\vec a$;
\bea
&&E_{\pm\vec b_{ij}}\ ,\quad E_{\vec a_{ijk}}\ ,\nn\\
&&\vec H' =\vec H -\ft58(\vec a\cdot\vec H)\, \vec a\ . 
\eea
On the other hand, the other negative-root generators $E_{-\vec
a_{ijk}}$ act on $\ket{-\vec a}$ to give $\ket{\vec a_\ell}$, where
$i,j,k$ and $\ell$ are all different.  The conjugate pairs $E_{\pm\vec
b_{ij}}$, together with the three Cartan generators $\vec H'$ ,
generate $SL(4,\R)$.  The four generators $E_{\vec a_{ijk}}$ mutually
commute, and form the vector representation under $SL(4,\R)$, with
weight vectors $\vec a_{ijk} +\ft54 \vec a$ under $\vec H'$.  Thus the
stability group of a single 3-form charge in $D=7$ is $SL(4,\R)\semi
\R^4$, and the coset is
\be
\fft{SL(5,\R)}{SL(4,\R)\semi \R^4}\ .\label{d73f}
\ee
The coset has dimension 5, which equals the dimension of the space of
3-form charges.  Indeed, any point in this 5-dimensional vector space
can be reached by acting on the original single charge vector with
$SL(5,\R)$.  As usual, this can be seen by observing that all the
states of the 5-dimensional representation for the 3-forms can be
reached by the action of the generators of $SL(5,\R)$ that do not
annihilate the highest-weight state $\ket{-\vec a}$.  The stability
group in (\ref{d73f}) can also be understood as follows: In the $D=7$
Lagrangian obtained by direct dimensional reduction from $D=11$
without any dualisation, the global symmetry group is $SL(4,\R)\semi
\R^4$ \cite{tdual,cjlp}, under which the 4-form is a singlet.

     Let us now consider the 2-forms in $D=7$.  The highest-weight
vector is $\vec a_{34}$.  It is easily verified that the following
subset of the $SL(5,\R)$ generators annihilate $\ket{\vec a_{34}}$:
\bea
&&E_{\pm\vec b_{12}}\ ,\quad E_{\pm\vec b_{34}}\ ,
\quad E_{\pm\vec a_{123}}\ ,\quad E_{\pm\vec a_{124}}\ ,\nn\\
&&\vec H'=\vec H - \ft5{12} (\vec a_{23}\cdot\vec H)\, \vec a_{23}\ ,
\label{d7gen1}\\
&&E_{\vec a_{134}}\ ,\quad E_{\vec a_{234}}\ ,\quad
E_{\vec b_{13}}\ ,\quad E_{\vec b_{23}}\ , \quad
E_{\vec b_{14}}\ ,\quad  E_{\vec b_{24}}\ .\nn
\eea
The conjugate pair $E_{\pm\vec b_{12}}$ together with the projection
of $\vec H'$ onto $\vec b_{12}$ generate an $SL(2,\R)$, while the
remaining three conjugate pairs, and the remaining two components of
$\vec H'$ generate an $SL(3,\R)$. The remaining six generators in
(\ref{d7gen1}) mutually commute, and form a (3,2) under the
$SL(3,\R)\times SL(2,\R)$; the highest weight generator is $E_{\vec
a_{234}}$, with weight vector $\vec a_{234}-\ft56\vec a_{34}$ under
$\vec H'$.  The coset parameterising the single-charge 2-form
solutions in $D=7$ is therefore
\be 
\fft{SL(5,\R)}{(SL(3,\R)\times SL(2,\R))\semi \R^6} \ .\label{d72f1}
\ee
The denominator group can also be easily understood from the viewpoint
of dimensional reduction from $D=8$ to $D=7$.  In $D=8$, the global
symmetry group is $SL(3,\R)\times SL(2,\R)$, which leaves the new
Kaluza-Klein vector invariant.  The $R^6$ comes from the commuting
shift symmetries of the six axions that are dimensional reduction of
six vectors in $D=8$, which form a $(3,2)$-dimensional representation
under $SL(3,\R)\times SL(2,\R)$.

      The coset (\ref{d72f1}) for a simple single-charge orbit is
7-dimensional, and so it does not parameterise all possible charge
points in the 10-dimensional vector space for 2-forms.  In the
infinitesimal neighbourhood of the charge vector for the field
strength $\F2{34}$, three charges, carried by field strengths
$\F2{12}$, $\cF2{3}$ and $\cF2{4}$, cannot be reached.  As discussed
in the introduction, when this happens we can construct a simple
2-charge solution, where the original charge is augmented by any of
the above three charges. To cover the entire space, we must then start
with a simple 2-charge configuration.  Any simple 2-charge solution
provides an equivalent starting point for generating the charge space,
since they form a single multiplet under the Weyl group of the global
symmetry group \cite{lpsweyl}.  Thus without loss of generality, we
shall consider a 2-charge solution constructed from the pair of field
strengths $\{\F2{12}, \F2{34}\}$, corresponding to the states
$\{\ket{\vec a_{12}}, \ket{\vec a_{34}}\}$.  It is easy to verify that
their sum is annihilated by the following $SL(5,\R)$ generators

\be
E_{\pm \vec b_{12}}\ ,\quad E_{\pm \vec b_{34}}\ ,\quad 
(E_{\pm \vec b_{24}} -E_{\mp\vec b_{13}})\ ,\quad
(E_{\pm\vec b_{14}} -E_{\mp\vec b_{23}})\ ,\quad
E_{\vec a_{ijk}}\ ,\label{d7hhh}
\ee
together with the two Cartan generators
\be
\vec H'=\vec H -\ft14(2\vec a_{34}\cdot \vec H + 3 \vec a_{12}\cdot 
\vec H) \, \vec a_{12} 
-\ft14(2\vec a_{12}\cdot \vec H + 3 \vec a_{34}\cdot 
\vec H) \, \vec a_{34}\ .
\ee
The conjugate pairs of generators in
(\ref{d7hhh}) have the weights
\be
\pm\vec b_{12}\ , \quad \pm\vec b_{34}\ ,\quad \pm\ft12(\vec
b_{24}-\vec b_{13})\ ,\quad \pm\ft12(\vec b_{14} -\vec b_{23})
\ee
under $\vec H'$.  We can now see that these are the roots of the
$B_2=O(2,3)$ algebra, where the simple roots can be taken to be
\be
\vec \a_1=b_{12}\ ,\quad \vec \a_2 = \ft12(\vec b_{24} - \vec b_{13})\ .
\ee
The generators $E_{\vec a_{ijk}}$ form a 4-dimensional
spinor representation under $O(2,3)$.  The highest-weight generator is
$E_{\vec a_{124}}$, which has weight $\ft12(\vec a_{124}-\vec
a_{123})$ under $\vec H'$.  Thus the coset of the $SL(5,\R)$
orbit of this pair of weights is given by 
\be
\fft{SL(5,\R)}{O(2,3)\semi \R^4}\ . \label{d72f2}
\ee 
This has dimensions 10, which is the same as the dimension of the
total charge space for 2-forms in $D=7$.  As usual, by looking at the
action of the $SL(5,\R)$ generators on the two states $\ket{\vec
a_{23}}$ and $\ket{\vec b_3}$, we can verify that indeed all generic
points in the 10-dimensional charge space can be reached.  The generic
ten-dimensional 2-charge orbits preserve $\ft14$ of the supersymmetry,
while the degenerate seven-dimensional single-charge orbits preserve
$\ft12$.

\section{$D=6$}

   The global the symmetry group in $D=6$ is $O(5,5)$, for which the
simple roots are $\vec b_{i,i+1}$, with $i=1,2,3,4$, and $\vec
a_{123}$.  The five 3-forms $\F3i$, together with their duals, form a
10-dimensional irreducible representation of $O(5,5)$, with weights
($\vec a_i, -\vec a_i$).  The highest-weight vector is $-\vec a_1$.
Splitting the internal index $i$, which runs over 5 values, as
$i=(1,\a)$, we easily see that the following subset of the $O(5,5)$
generators annihilate the state $\ket{-\vec a_1}$:
\bea
&&E_{\pm\vec b_{\a\b}}\ ,\quad E_{\pm \vec a_{1\a\b}}\ ,\quad E_{\vec
b_{1\a}}\ , \quad E_{\vec a_{\a\b\gamma}}\ ,\nn\\
&&\vec H'=\vec H -\ft12 (\vec a_1\cdot\vec H)\, \vec a_1\ ,\label{d6hhh}
\eea
while the remaining generators give
\be
E_{-\vec b_{1\a}}\, \ket{-\vec a_1} = \ket{-\vec a_\a}\ ,\quad
E_{-\vec a_{\a\b\gamma}}\, \ket{-\vec a_1}=\ket{\vec a_\delta}\ .
\ee
where in the last case the indices $\a,\b,\gamma$ and $\delta$, which
take their values in the set (2,3,4,5), are all different.  
The conjugate pairs of generators in (\ref{d6hhh}) have weights
\be
\pm\vec b_{\a\b}\ ,\quad \pm \vec a_{1\a\b}\ ,\quad \vec b_{1\a}\ , \quad
\vec a_{\a\b\gamma}\ ,
\ee
under $\vec H'$, and they are precisely the root vectors of
$D_4=O(4,4)$.  The simple roots may be taken to be
\be
\vec \a_1=\vec b_{23}\ ,\quad \vec \a_2=\vec b_{34}\ ,
\quad\vec \a_3=\vec b_{45}\ ,
\quad \vec\a_4=\vec a_{123}\ .  
\ee
The remaining eight generators $E_{\vec b_{1\a}}$ and $E_{\vec
a_{\a\b\gamma}}$ in (\ref{d6hhh}) form an 8-dimensional representation
under $O(4,4)$.  The highest-weight state is $E_{\vec a_{345}}$, with
weight $\vec a_{345}+\vec a_1 =-\vec a_2$ under $\vec H'$.
Thus the coset parameterising the single-charge 3-form solutions in
$D=6$ is
\be
\fft{O5,5)}{O(4,4)\semi \R^8}\ .\label{d63f1}
\ee
There is again an easier explanation for the denominator
group that leaves a single 3-form charge invariant.  In $D=6$, there
is an $O(4,4)$ T-duality, under which the NS-NS 3-form field strength
$\F31$ is a singlet.  In addition, the eight R-R axions 
$\cF1{1\a}$, $\F1{\a\b\gamma}$, which can all be covered by  
derivatives simultaneously, form an 8-dimensional spinor representation
under the T-duality group.  In fact $O(4,4)\semi \R^8$ is precisely
the global symmetry of the version of the supergravity obtained by
dualising only the higher-degree R-R fields \cite{tdual}. 

    The coset (\ref{d63f1}) has dimension 9, which is one less than
the dimension of the charge-vector space for 3-forms.  In fact, the
denominator group in (\ref{d63f1}) is the stability group of a
light-like vector in $O(5,5)$.  We shall return to this point below.
For now, let us proceed by extending the discussion to 2-charge
solutions.  These are in fact dyonic (of the first kind, in the
notation of \cite{lpsol}), where a single field strength carries both
electric and magnetic charges \cite{dfkr}.  To study the 2-charge
orbits, we may therefore consider the stability group for the pair of
states $\ket{\vec a_1}$ and $\ket{-\vec a_1}$.  It is easily
established that their sum is left invariant by the subset of $O(5,5)$
generators 
\be
E_{\pm\vec b_{\a\b}}\ ,\quad E_{\pm\vec a_{1\a\b}}\ , \quad 
(E_{\pm\vec a_{\a\b\gamma}}
-E_{\mp\vec b_{1\delta}})\ ,\label{d6rootgen}
\ee
where in the last case $\a,\b,\gamma$ and $\delta$ are all different,
together with the same set of four Cartan generators $\vec H'$ given
in (\ref{d6hhh}). (Since the weights $\vec a_1$ and $-\vec a_1$ are 
parallel, there are still four combinations of the $O(5,5)$ Cartan 
generators that leave the two states invariant.)  The generators in 
(\ref{d6rootgen}) have weights 
\be
\pm\vec b_{\a\b}\ ,\quad \pm\vec a_{1\a\b}\ , \quad\pm 
\ft12(\vec a_{\a\b\gamma}
-\vec b_{1\delta}) \label{d6root}
\ee
under $\vec H'$; these are precisely the roots of $B_4=O(4,5)$.  
The simple roots may be taken to be
\be
\vec \a_1=\vec b_{23}\ , \quad \vec \a_2=\vec b_{34}\ ,\quad  
\vec\a_3= \vec b_{45}\ ,\quad
\vec\a_4=\ft12(\vec a_{234} -\vec b_{15})\ .
\ee
Thus the coset parameterising the 2-charge 3-form solutions is
\be
\fft{O(5,5)}{O(4,5)}\ .\label{d63f2}
\ee
Note that in this case all the generators that annihilate the states
lie in the single simple group $O(4,5)$, and there are no additional
$\R$ factors.  

     The denominator group in the 2-charge coset is the stability
group for a timelike or spacelike vector in $O(5,5)$, and indeed the
coset (\ref{d63f2}) has dimension 9. This is what one expects for
timelike or spacelike 10-vectors of fixed length.  The rescaling of
the length can be implemented by using the trombone symmetry of the
$D=6$ equations of motion.  By including this rescaling, all points in
the 10-dimensional charge space for 3-forms in $D=6$ correspond to
valid string solutions.  There are three independent orbits; timelike,
spacelike and null.  To understand this, we note that the electric
charges $Q_i$ and the magnetic charges $P_i$ assemble into the vector
representation of $O(5,5)$ in the form $(Q_i+P_i,Q_i-P_i)$.  The
magnitude of this vector is
\be
\sum_i(Q_i+P_i)^2 -\sum_i(Q_i-P_i)^2 = 4\sum_i Q_i\, P_i\ .
\ee
Thus we see that the charge vector is null on the orbit of any
single-charge solution, and timelike or spacelike on any of the
2-charge orbits.  Note that the null orbits preserve $\ft12$ the
supersymmetry, while the timelike and spacelike orbits preserve
$\ft14$. It is worth remarking that the self-dual string and the
anti-self-dual string belong to different multiplets under $O(5,5)$,
with the former lying on the spacelike orbit, and the latter on the
timelike orbit.
 
      Another way to understand why the dyonic strings with different
relative signs of their electric and magnetic charges belong to
different multiplets is the following.  The weight vectors for the
electric and magnetic charges are anti-parallel, and hence there is
only a single Cartan generator that can rescale the charges, and
therefore it cannot change the relative sign of the two charges.  As
we shall see in the next two sections, the weight vectors for simple
3-charge solutions in $D=5$ and for simple 4-charge solutions in $D=4$
are again degenerate, in that they lie in 2-dimensional and
3-dimensional planes respectively.  Thus the number of Cartan
generators that can rescale the charges is one less than the number of
charges.  It follows that in both these cases, configurations with
different relative signs of the charges belong to different
multiplets.  It is interesting to note that this phenomenon always
arises in the cases where the dilatonic scalar fields in the solution
are regular, even at the horizon.
       
     In the fully-dualised six-dimensional theory there are in total
sixteen 2-forms, namely $\F2{ij}$, $\cF2i$ and $*F_4$.  Their dilaton
vectors $\vec a_{ij}$, $\vec b_i$ and $ -\vec a$ form an irreducible
16-dimensional representation of $O(5,5)$, with $-\vec a$ as the
highest weight.  The state $\ket{-\vec a}$ is annihilated by the
following subset of the $O(5,5)$ generators:
\bea
&&E_{\pm \vec b_{ij}}\ ,\quad E_{ \vec a_{ijk}}\ ,\nn\\
&&\vec H'=\vec H -\ft25 (\vec a\cdot\vec H)\, \vec a\ ,\label{d6hhhh}
\eea
while the generators $E_{-\vec a_{ijk}}$ give
\be
E_{-\vec a_{ijk}}\, \ket{-\vec a} = \ket{\vec a_{\ell m}}\ ,
\ee
where $i,j,k,\ell$ and $m$ are all different.  The conjugate pairs of
generators in (\ref{d6hhhh}) have weights $\pm \vec b_{ij}$ under 
$\vec H'$; these are the roots of the algebra
$SL(5,\R)$, for which the simple roots may be taken to be
\be
\vec\a_1=\vec b_{12}\ ,\quad \vec\a_2=\vec b_{23}\ ,\quad 
\vec\a_3=\vec b_{34}\ ,\quad \vec\a_4=\vec b_{45}\ .
\ee
The remaining generators $E_{\vec a_{ijk}}$ in (\ref{d6hhhh}) mutually
commute, and form a 10-dimensional representation under $SL(5,\R)$. 
The highest-weight generator is $E_{\vec a_{345}}$, which has weight
$\vec a_{345}+\ft45\vec a$ under $\vec H'$. 
Therefore, the coset parameterising the single-charge 2-form solutions
in $D=6$ is
\be
\fft{O(5,5)}{SL(5,\R)\semi \R^{10}}\ .\label{d62f1}
\ee
Note that the $SL(5,\R)$ can be understood as being inherited directly
from the global symmetry group in $D=7$.  In particular, if we were to
consider a single-charge solution in $D=6$ using the new Kaluza-Klein
2-form $\cF25$ rather than $*F_4$, then the $SL(5,\R)$ of $D=7$ would
leave it invariant.  Also, the constant shift symmetries $\R^{10}$ of
the ten axions coming from the dimensional reduction of the ten
vectors in $D=7$ also leave the new Kaluza-Klein vector invariant.
Since the solution using $\cF25$ in $D=6$ is equivalent, under
$O(5,5)$, to the solution using $*F_4$, the stability group will be
the same (modulo conjugations) in each case.

     The dimension of the coset (\ref{d62f1}) for single-charge 2-form
solutions in $D=6$ is 11, while the total charge-space for 2-forms has
dimension 16.  General points in the charge space can be reached by
looking at the orbits of 2-charge solutions.  We shall start from the
simple 2-charge solution supported by $*F_4$ and $\cF25$,
corresponding to the sum of the states $\ket{-\vec a}$ and $\ket{\vec
b_5}$.  We find that this sum of states is annihilated by the
following subset of $O(5,5)$ generators:
\bea
&&E_{\pm \vec b_{ab}}\ ,\quad (E_{\pm\vec a_{ab5}} -E_{\mp\vec
a_{cd5}})\ ,\quad
E_{-\vec b_{a5}}\ , \quad E_{\vec a_{abc}}\ ,\label{d6root2}\\
&&\vec H'=\vec H +\ft18(3\vec a\cdot\vec H -5 \vec b_5\cdot\vec H)\,
\vec b_5 +\ft18(3\vec b_5\cdot\vec H -5\vec a\cdot\vec H)\, \vec a
\eea
where we have split $i=(a,5)$, and in the second case $a,b,c$ and $d$
are all different.  The conjugate pairs of generators have weights
\be
\pm \vec b_{ab}\ ,\quad \pm\ft12 (\vec a_{ab5} -\vec a_{cd5})\label{d6root3}
\ee
under $\vec H'$; these are the roots of the algebra $B_3=O(3,4)$.  The
simple roots may be taken to be
\be
\vec\a_1=\vec b_{12}\ ,\quad \vec\a_2=\vec b_{23}\ ,\quad 
\vec \a_3=\ft12(\vec a_{145} -\vec a_{235})\ .
\ee
The further eight generators $E_{-\vec b_{a5}}$ and $E_{\vec a_{abc}}$ in
(\ref{d6root2}) mutually commute, and form a
spinor representation under $O(3,4)$.  The highest-weight generator is
$E_{-\vec b_{45}}$, which has weight $-\vec b_{45} -\ft54\vec b_5
+\ft34\vec a$.  The coset parameterising the
2-charge 2-form solutions in $D=6$ is therefore
\be
\fft{O(5,5)}{O(3,4)\semi \R^8}\ .\label{d62f2}
\ee
This has dimension 16, which is the same as the total dimension of the
charge space for 2-forms in $D=6$.  Indeed, one can reach all generic
points in the charge vector space, by starting from a simple 2-charge.
The degenerate eleven-dimensional single-charge orbits correspond as
usual to solutions that preserve $\ft12$ the supersymmetry, while the
generic sixteen-dimensional 2-charge orbit preserves $\ft14$.

\section{$D=5$}

     The global symmetry group for the fully-dualised theory in $D=5$
is $E_6$.  The simple roots are $\vec b_{i,i+1}$, with $i=1, 2,3,4,5$,
and $\vec a_{123}$.  The full set of roots is $\pm\vec b_{ij}$, $\pm
\vec a_{ijk}$, and $\mp\vec a$.  In the fully-dualised theory, there
are only 2-form field strengths (in addition to the axions), namely
$\{\F2{ij},\cF2i, *\F3i\}$.  Their dilaton vectors $\{\vec a_{ij},
\vec b_i,-\vec a_i\}$ are the weights of the 27-dimensional
irreducible representation of $E_6$, with $-\vec a_1$ as the highest
weight.  The associated state $\ket{-\vec a_1}$ is annihilated by the
subset of $E_6$ generators
\bea
&&E_{\pm \vec b_{\a\b}}\ ,\quad E_{\pm \vec a_{1\a\b}}\ , \quad
E_{\vec b_{1\a}}\ ,\quad E_{\vec a_{\a\b\gamma}}\ ,\quad 
E_{-\vec a}\ ,\label{d5xx1gen}\\
&&\vec H'=\vec H -\ft38 (\vec a_1\cdot\vec H)\, \vec a_1\ ,
\eea
where we have split $i=(1,\a)$.  The conjugate pairs of generators
have weights
\be
\pm \vec b_{\a\b}\ ,\quad \pm a_{1\a\b} \label{d5xx1}
\ee
under $\vec H'$, and thus form the roots of the algebra $D_5=O(5,5)$.
The simple roots can be taken to be
\be
\vec\a_1= \vec b_{23}\ ,\quad \vec\a_2=\vec b_{34}\ ,
\quad \vec\a_3=\vec b_{45}
\ ,\quad \vec\a_4=\vec b_{56}\ ,\quad \vec\a_5=\vec a_{123}\ .
\ee
The remaining generators $E_{\vec b_{1\a}}$, $E_{\vec a_{\a\b\gamma}}$
and $E_{-\vec a}$ in
(\ref{d5xx1gen}) are mutually-commuting, and form a
16-dimensional spinor representation of $O(5,5)$.  The highest-weight 
generator is $E_{-\vec a}$, with weight $-\vec a+\ft34 \vec a_1$ under
the $O(5,5)$ Cartan generators $\vec H'$.  The coset
parameterising the single-charge 2-form solutions in $D=5$ is
therefore
\be
\fft{E_6}{O(5,5)\semi \R^{16}}\ .\label{d52f1}
\ee
There are two more intuitive ways to understand the denominator group.
One is to observe that the NS-NS 3-form $\F31$ is a singlet under the
T-duality group $O(5,5)$, and it is also invariant under the $\R^{16}$
shift symmetries of the sixteen R-R axions
$\{\cF1{1\a},\F1{\a\b\gamma}, *F_4\}$.  Another way to understand the
denominator group is to view it as being inherited from $D=6$, which
obviously leaves the new Kaluza-Klein vector $\cF26$ invariant.  The
$\R^{16}$ in this case is generated by the commuting shift symmetries
of the axions coming from the dimensional reduction of the sixteen
2-forms in $D=6$.  Since the new Kaluza-Klein vector is equivalent,
under $E_6$, to the 2-form $*\F31$ that we were previously
considering, it follows that the stability group will also be the
same, modulo conjugations.

   Moving now to 2-charge 2-form solutions in $D=5$, we may take these
to be represented by the states $\ket{-\vec a_1}$ and $\ket{\vec
a_{16}}$.  This combination is left invariant by the subset of $E_6$ 
generators
\bea
&&E_{\pm\vec b_{ab}}\ ,\quad E_{\pm \vec a_{1ab}}\ ,\quad (E_{\pm\vec
a_{abc}}-E_{\mp\vec b_{1d}})\ ,\nonumber \\
&& E_{\vec b_{16}}\ ,\quad E_{\vec b_{a6}}\ ,\quad E_{\vec a_{1a6}}\ ,
\quad E_{\vec a_{ab6}}\ ,\quad
E_{-\vec a}\ ,\label{d5xx2gen}\\
&&\vec H'=\vec H +\ft14(\vec a_1\cdot \vec H -2\vec a_{16}\cdot \vec
H) \, \vec a_{16} +\ft14(\vec a_{16}\cdot\vec H -2 \vec a_1\cdot\vec
H)\, \vec a_1\ ,\nn
\eea
where we have split the internal indices as $i=(1,a,6)$, and in the
final expression on the top line, $a$, $b$, $c$ and $d$ are all
different.  We find that the conjugate pairs of generators have weights
\bea
&&\pm\vec b_{ab}\ ,\quad \pm \vec a_{1ab}\ ,\quad \pm \ft12(\vec
a_{abc}-\vec b_{1d}) \label{d5xx2}
\eea
under $\vec H'$.  These are the roots of the algebra $B_4=O(4,5)$, for
which we may take the simple roots to be
\be
\vec \a_1 = \vec b_{23}\ ,\quad \vec \a_2=\vec b_{34}\ ,\quad \vec
\a_3=\vec b_{45}\ , \quad \vec a_4=\ft12(\vec a_{234}-\vec b_{15})\ .
\ee
The remaining generators in 
(\ref{d5xx2gen}) mutually commute, and form a 16-dimensional
irreducible spinor representation of $O(4,5)$.  The highest weight
generator is $E_{-\vec a}$, with weight $-\ft12(\vec b_{16}+\vec a)$
under $\vec H'$.  The coset parameterising the
2-charge 2-form solutions in $D=5$ is therefore
\be
\fft{E_6}{O(4,5)\semi \R^{16}}\ .\label{d52f2}
\ee
The denominator group can in fact also be understood from the
observation that the 2-charge solution using $\{*\F31,\F2{16}\}$ is
the dimensional reduction of the dyonic string in $D=6$, supported by
electric and magnetic charges for $\F31$.  This field strength was
already previously found to be invariant under $O(4,5)$.  The
commuting $\R^{16}$ symmetries are the shift symmetries of the sixteen
axions coming from the dimensional reduction of the sixteen 2-forms in
$D=6$.

     The dimension of the coset (\ref{d52f2}) for 2-charge 2-forms in
$D=5$ is 26, which is one less than the total dimension of the charge
vector space for 2-forms.  Thus we see that not all points in the
vector space can be reached as the orbits of a simple 2-charge
solution.  In particular, starting from the weight vectors $-\vec a_1$
and $\vec a_{16}$, we cannot generate the weight vector $\vec b_6$.
However, there exist simple 3-charge solutions in $D=5$, precisely
using the field strengths $\{*\F31,\F2{16}, \cF26\}$ \cite{lpsol}.
Now, we may therefore augment the previous discussion by considering
as a starting point the 3-charge configuration represented as the sum
of states $\ket{-\vec a_1}$, $\ket{\vec a_{16}}$ and $\ket{\vec b_6}$.
It is easy to check as usual from the dilaton vector sum rules that
this combination of states is annihilated by the subset of $E_6$
generators 
\bea
&&E_{\pm\vec b_{ab}}\ ,\quad E_{\pm \vec a_{1ab}}\ ,\quad 
(E_{\pm\vec b_{1a}} -E_{\mp\vec a_{bcd}})\ ,\nonumber \\
&&(E_{\pm\vec a_{ab6}}-E_{\mp\vec a_{cd6}})\ ,\quad (E_{\pm\vec
b_{a6}}-E_{\mp\vec a_{1a6}})\ ,\quad (E_{\pm\vec b_{16}}- E_{\mp\vec a})\ ,
\label{d5xx7gen}
\eea
where as before we have split the internal indices as $i=(1,a,6)$,
together with the same subset of Cartan generators $\vec H'$ that is defined
in (\ref{d5xx2gen}).  (The reason why there are still four combinations
of the $E_6$ Cartan generators that annihilate the three states in
this 3-charge case is that the third charge has a dilaton vector that
is coplanar with the first two.)  It is straightforward to
verify that the full set of generators in (\ref{d5xx7gen}) have
the weights
\bea
&&\pm\vec b_{ab}\ ,\quad \pm \vec a_{1ab}\ ,\quad \pm \ft12 (\vec b_{1a}
-\vec a_{bcd})\ ,\nonumber \\
&&\pm\ft12(\vec a_{ab6}-\vec a_{cd6})\ ,\quad \pm\ft12(\vec
b_{a6}-\vec a_{1a6})\ ,\quad \pm \ft12(\vec b_{16} +\vec a)
\eea
under $\vec H'$, and that they are nothing but the roots of the $F_4$ 
algebra.  The simple roots can be taken to be
\be
\vec\a_1 = \vec b_{34}\ ,\quad \vec a_2=\vec b_{45}\ ,\quad \vec
\a_3=\ft12(\vec a_{234}-\vec b_{15})\ ,\quad
\vec\a_4=\ft12(\vec b_{16} +\vec a)\ .
\ee
The coset parameterising the 3-charge 2-form solutions in $D=5$ is
therefore
\be
\fft{E_6}{F_4}\ .\label{d52f3}
\ee
The fact that the denominator group $F_4$ contains $O(4,5)$ as a
subgroup can be understood from the fact that the 3-charge solution
becomes a boosted dyonic string, \ie an intersection of a pp wave and
a dyonic string, in $D=6$, whose charges are obviously invariant under the
$O(4,5)$ stability group that we found in section 5.  

     The dimension of the 3-charge coset space (\ref{d52f3}) is 26, as
is the dimension of the 2-charge coset space (\ref{d52f2}).  In fact
the 2-charge orbits correspond to null vectors in the 27-dimensional
charge vector space, while the 3-charge orbits correspond to timelike
or spacelike vectors \cite{fg}.  The reason why these 3-charge orbits
have only dimension 26 is that an overall scaling symmetry, \ie the
trombone symmetry, is also needed in order to fill out the entire
27-dimensional charge vector space.  This can be seen from the fact
that the three weight vectors $\{-\vec a_1,\vec a_{16},\vec b_6\}$
associated with the three charges are coplanar, and so only two,
rather than three, $E_6$ Cartan generator combinations are able to
scale the charges.  This is analogous to the situation that we saw for
dyonic strings in $D=6$.  This situation contrasts with that for the
26-dimensional orbits for the 2-charge solutions in $D=5$, where, as
we saw above it is a non-zero weight, \ie $\vec b_6$, rather than a
scaling of one of the basic charges, whose absence reduces the
dimension of the orbits from 27 to 26.

    The 3-charge orbits preserve $\ft18$ of the supersymmetry, while
the 2-charge orbits preserve $\ft14$ and the 1-charge orbits preserve
$\ft12$.

\section{$D=4$}
 
    The global symmetry group of the fully-dualised theory in $D=4$ is
$E_7$.  The simple roots are $\vec b_{i,i+1}$ with $1\le i\le 6$, and
$\vec a_{123}$.  The full set of roots is $\pm\vec b_{ij}$, $\pm \vec
a_{ijk}$ and $\mp\vec a_i$.  There are 28 2-forms $\F2{ij}$ and
$\cF2i$ which, together with their duals, form a 56-dimensional
irreducible representation of $E_7$.  The associated dilaton vectors
$\{\vec a_{ij}, \vec a_i, -\vec a_{ij}, -\vec b_i\}$ are the weights
of the 56.  The highest weight vector is $-\vec b_7$.  We find that
the associated state $\ket{-\vec b_7}$ is annihilated by the following
subset of the $E_7$ generators:
\bea
&&E_{\pm b_{ab}}\ ,\quad E_{\pm \vec a_{abc}}\ ,\quad E_{\mp \vec a_7}\ , 
\quad E_{\vec b_{a7}}\ , \quad E_{\vec a_{ab7}}\ ,
\quad E_{-\vec a_a}\ ,\nn\\
&&\vec H'=\vec H -\ft13 (\vec b_7\cdot\vec H)\, \vec b_7
\ ,\label{d4xx1gen}
\eea
where we have split the internal index as $i=(a,7)$.  We find that the
weights of the conjugate pairs of generators under $\vec H'$ are given by
\be
\pm b_{ab}\ ,\quad \pm \vec a_{abc}\ ,\quad \mp \vec a_7\ , 
\ ,\label{d4xx1}
\ee
and that these are precisely the roots of the $E_6$ algebra. 
The simple roots may be taken to be
\be
\vec \a_1=\vec b_{12}\ ,\quad \vec \a_2 = \vec b_{23}\ ,\quad
\vec \a_3 = \vec b_{34}\ ,\quad \vec \a_4 = \vec b_{45} ,\quad
\vec \a_5= \vec b_{56}\ ,\quad \vec \a_6=\vec a_{123}\ .
\ee
The further 27 generators in (\ref{d4xx1gen}) mutually commute, and
form an irreducible 27-dimensional representation under $E_6$, with
$E_{-\vec a_1}$ as the highest-weight generator.   It
follows that the coset that parameterises the single-charge 2-form
solutions in $D=4$ is
\be
\fft{E_7}{E_6\semi \R^{27}}\ .\label{d42f1}
\ee
The denominator group can be easily understood since $E_6$ is
inherited directly from $D=5$, and the $\R^{27}$ symmetry is generated
by the 27 axions that are the dimensional reductions of the 27 2-form
field strengths in $D=5$.  The coset (\ref{d42f1}) has dimension 28.
Acting on a single-charge solution with $E_7$ gives rise to a solution
which can carry a maximum of 28 charges.  In order to fill out more
points in the 56-dimensional charge space, we need to look at orbits
involving multi-charge simple solutions.

         Simple 2-charge solutions can be constructed using the field
strength $\{*\cF27, *\F2{17}\}$, associated with the dilaton vectors
$\{-\vec b_7, - \vec a_{17}\}$.  We find that the subset of the $E_7$ 
generators that leave the combination of states $\ket{-\vec b_7}$ and 
$\ket{-\vec a_{17}}$ invariant is
\bea
&&E_{\pm \vec b_{ab}}\ ,\quad E_{\pm \vec a_{1ab}}\ ,\quad
(E_{\pm\vec b_{a7}} - E_{\mp\vec a_{1a7}})\ ,\nn\\
&&E_{\vec a_{abc}}\ ,\quad E_{\vec a_{ab7}}\ ,\quad
E_{-\vec a_1}\ ,\quad E_{-\vec a_a}\ ,\quad E_{-\vec a_7}
\ ,\quad E_{\vec b_{1a}}\ ,\quad E_{\vec b_{17}}\ ,\label{d4xx2gen}\\
&&\vec H'=\vec H +\ft18(\vec b_7\cdot\vec H -3\vec a_{17}\cdot\vec H)\,
\vec a_{17} +\ft18(\vec a_{17}\cdot\vec H -3 \vec b_7\cdot\vec H)\, 
\vec b_7\ ,
\eea
where we have split the index $i=(1,a,7)$. We find that
the conjugate pairs of generators in (\ref{d4xx2gen}) have weights 
\be
\pm \vec b_{ab}\ ,\quad \pm \vec a_{1ab}\ ,\quad
\pm \ft12 (\vec b_{a7} - \vec a_{1a7})\ ,\label{d4xx2}
\ee
under $H'$, and that these are precisely the roots of the 
$B_5=O(5,6)$ algebra.  The simple roots can be taken to be
\be
\vec\a_1=\vec b_{23}\ ,\quad \vec \a_2 = \vec b_{34}\ ,\quad \vec \a_3 = 
\vec b_{45}\ , \quad \vec \a_4= \vec b_{56}\ ,\quad \vec \a_5=
\ft12(\vec b_{67}-\vec a_{167})\ .
\ee
The remaining generators in (\ref{d4xx2gen}), of which there are 33 in
total, form a 32-dimensional spinor representation under $O(5,6)$ together
with a singlet, corresponding to the generator $E_{-\vec a_1}$.   The
highest-weight generator in the 32 is $E_{-\vec a_7}$, which has
weight $-\vec a_7 +\ft34\vec a_{17} +\ft14 \vec b_7$.  Thus the
coset for 2-charge 2-form solutions in $D=4$ is
\be
\fft{E_7}{(O(5,6)\semi \R^{32})\times \R}\ .\label{d42f2}
\ee
This coset has dimension 45.

     For 3-charge solutions, we may use the field strengths 
$\{*\cF27, *\F2{17}, \F2{16}\}$, corresponding to the weight vectors 
$\{-\vec b_7, - \vec a_{17}, \vec a_{16}\}$.  We find that the subset
of $E_7$ generators that annihilate this combination of three states
is given by
\bea
&&E_{\pm \vec b_{ab}}\ ,\quad E_{\pm \vec a_{1ab}}\ ,\quad (E_{\pm\vec b_{1a}}
-E_{\mp\vec a_{bcd}})\ ,\quad (E_{\pm\vec b_{17}} -E_{\pm\vec a_6})\ ,\quad
(E_{\pm\vec b_{a7}}-E_{\mp\vec a_{1a7}})\ ,\nn\\
&&(E_{\pm\vec a_{ab7}}-E_{\mp\vec a_{cd7}})\ ,\quad E_{\vec b_{16}}\ ,\quad
E_{\vec b_{a6}}\ ,\quad E_{\vec a_{1a6}}\ ,\quad E_{\vec a_{ab6}}\ ,\nn\\
&& E_{\vec a_{a67}}\ ,\quad E_{-\vec a_a}\ ,\quad E_{-\vec a_7} \ ,\quad
(E_{\vec a_{167}}-E_{-\vec b_{67}})\ , \quad (E_{-\vec a_1}-E_{-\vec b_{67}})
\ ,\label{d4xx3gen}\\
&&\vec H' =\vec H +\ft14(\vec a_{16}-\vec a_{17}-2\vec b_7)\cdot \vec
H \, \vec b_7 +\ft14(\vec a_{16}-2\vec a_{17}-\vec b_7)\cdot \vec
H \, \vec a_{17} \nn\\
&&\qquad\qquad +\ft14(-2\vec a_{16}+\vec a_{17}+\vec b_7)\cdot \vec
H \, \vec a_{16}\ ,\nn
\eea
where we have split the internal indices as $i=(1,a,6,7)$.  We find
that the weights of the conjugate pairs of generators in
(\ref{d4xx3gen}) under $\vec H'$ are
\bea
&&\pm \vec b_{ab}\ ,\quad \pm \vec a_{1ab}\ ,\quad \pm\ft12(\vec b_{1a}
-\vec a_{bcd})\ ,\quad \pm\ft12(\vec b_{17}+\vec a_6)\ ,\nn \\
&&\pm\ft12(\vec b_{a7}-\vec a_{1a7})\ ,\quad
\pm\ft12 (\vec a_{ab7}-\vec a_{cd7})\ , \label{d4xx3}
\eea
and that these coincide with the roots of the $F_4$ algebra.
The simple roots may be taken to be
\be
\vec\a_1=\vec b_{34}\ ,\quad \vec\a_2=\vec b_{45}\ ,\quad
\vec\a_3=\ft12(\vec a_{234}-\vec b_{15})\ ,\quad 
\vec\a_4=\ft12(\vec b_{17}+\vec a_6)\ .
\ee
The remaining generators in (\ref{d4xx3gen}) form a 26-dimensional
representation under $F_4$, with $E_{\vec a_{267}}$ as the
highest-weight generator, with weight $\ft12(\vec a_{267}+\vec a_2)$.
Thus the coset parameterising the 3-charge 2-form solutions in $D=4$ is
\be
\fft{E_7}{F_4\semi \R^{26}}\ .\label{d42f3}
\ee
Note that the last two generators listed in (\ref{d4xx3gen}) have zero
weight under the Cartan generators $\vec H'$ of $F_4$; this can be
seen from the fact that these two vectors have components only in the
plane orthogonal to the 4-dimensional space spanned by the $F_4$
roots.  In fact, they are the two zero weights in the 26-dimensional
representation of $F_4$.

     The dimension of the coset (\ref{d42f3}) is 55, which is one less
than the dimension of the charge vector space for 2-forms in $D=4$.
In particular, if we start with the three charges corresponding to the
weight vectors $\{-\vec b_7,-\vec a_{17},\vec a_{16}\}$, we can never
generate the charge associated with the weight vector $\vec b_6$ by
acting with the $E_7$ global symmetry group.  In fact there exists a
simple 4-charge solution that uses precisely this extra charge in
addition to the previous three.  Thus we may now take this 4-charge
configuration as a new starting point, and examine its orbit under
$E_7$.  We find that the corresponding combination of four states is
annihilated by the following subset of $E_7$ generators
\bea
&&E_{\pm\vec b_{ab}}\ ,\quad E_{\pm \vec a_{1ab}}\ ,\quad 
(E_{\pm\vec b_{1a}} -E_{\mp\vec a_{bcd}})\ ,\nonumber \\
&&(E_{\pm\vec a_{ab6}}-E_{\mp\vec a_{cd6}})\ ,\quad (E_{\pm\vec
b_{a6}}-E_{\mp\vec a_{1a6}})\ ,\quad 
(E_{\pm\vec b_{16}} -E_{\pm\vec a_7})\ ,\label{d4xx5gen}\\
&&\vec H' =\vec H +\ft14(\vec a_{16}-\vec a_{17}-2\vec b_7)\cdot \vec
H \, \vec b_7 +\ft14(\vec a_{16}-2\vec a_{17}-\vec b_7)\cdot \vec
H \, \vec a_{17} \nn\\
&&\qquad\qquad +\ft14(-2\vec a_{16}+\vec a_{17}+\vec b_7)\cdot \vec
H \, \vec a_{16} \nn
\eea
and 
\bea
\!\!\!\!\! &&(E_{\pm\vec b_{17}}-E_{\pm\vec a_6})\ ,
\quad (E_{\pm\vec b_{a7}}-E_{\mp\vec
a_{1a7}})\ ,\quad (E_{\pm\vec a_{ab7}}-E_{\mp\vec a_{cd7}})\ ,\quad
(E_{\pm\vec a_{a67}} -E_{\pm\vec a_a})\ ,\label{d4xx6gen}\\
\!\!\!\!\! &&H_5=E_{\vec b_{67}}+ E_{-\vec b_{67}}-E_{\vec a_{167}} 
-E_{-\vec a_{167}}\ ,\quad
H_6=E_{\vec b_{67}}+ E_{-\vec b_{67}}-E_{\vec a_{1}} 
-E_{-\vec a_{1}}\ ,\label{extra}
\eea
where we have again split the internal index as $i=(1,a,6,7)$.  It is
easy to see that the generators in (\ref{d4xx5gen}) form an $F_4$
algebra, for which the non-zero roots, measured by $\vec H'$, are
\bea
&&\pm\vec b_{ab}\ ,\quad \pm \vec a_{1ab}\ ,\quad \pm \ft12 (\vec b_{1a}
-\vec a_{bcd})\ ,\nonumber \\
&&\pm\ft12(\vec a_{ab6}-\vec a_{cd6})\ ,\quad \pm\ft12(\vec
b_{a6}-\vec a_{1a6})\ ,\quad \pm \ft12(\vec b_{16} +\vec a_7)\ .\label{d4xx5}
\eea
The simple roots may be taken to be
\be
\vec\a_1 = \vec b_{34}\ ,\quad \vec a_2=\vec b_{45}\ ,\quad \vec
\a_3=\ft12(\vec a_{234}-\vec b_{15})\ ,\quad
\vec\a_4=\ft12(\vec b_{16}+\vec a_7)\ .\label{f4d4}
\ee
The 26 generators in (\ref{d4xx6gen}) and (\ref{extra}) form an
irreducible 26-dimensional representation under the $F_4$.  In
particular, $H_5$ and $H_6$ are the two states with zero weights under
$\vec H'$.  The weights of the remaining 24 generators 
(\ref{d4xx6gen}) under $\vec H'$ are given by
\be
\pm\ft12(\vec b_{17}+\vec a_6)\ ,\quad \pm\ft12(\vec b_{a7}-\vec
a_{1a7})\ ,\quad \pm\ft12(\vec a_{ab7}-\vec a_{cd7})\ ,\quad
\pm \ft12(\vec a_{a67}+\vec a_a)\ .\label{d4xx6}
\ee
In fact the full set of all the generators in (\ref{d4xx5gen}),
(\ref{d4xx6gen}) and (\ref{extra}) generate an $E_6$ algebra, with
$H_5$ and $H_6$ giving the extra two Cartan generators over and above
the four Cartan generators $\vec H'$ of $F_4$.  Thus we have a
description of $E_6$ in terms of its decomposition under $F_4$, with
$78\rightarrow 52+26$.  Note that the extra Cartan generators $H_5$
and $H_6$ are compact, unlike the standard non-compact Cartan
generators that have arisen previously, and so the version of $E_6$
that we obtain here is $E_{6(2)}$ rather than the maximally
non-compact form $E_{6(6)}$. Thus the coset parameterising the 4-charge
2-form solutions in $D=4$ is
\be
\fft{E_7}{E_{6(2)}}\ .\label{d42f4}
\ee
This result was also obtained in \cite{fg}.

     Before discussing the interpretation of these results for BPS
2-form solutions, we should first note that there is a new feature in
$D=4$, namely that there also exists another class of charge
configuration, which is associated with extremal solutions that are
not supersymmetric. Specifically, there exists a 2-charge dyonic
solution, where the
electric and magnetic charges are carried by the same field strength.
Thus, for example, we may consider the electric and magnetic charges 
associated with the
field strength $\cF27$.  We should therefore look for the stability
subgroup of $E_7$ that leaves the combination of states $\ket{-\vec
b_7}$ and $\ket{\vec b_7}$ invariant.  We find that the following
subset of the $E_7$ generators annihilate this combination of states:
\bea
&&E_{\pm\vec b_{ab}}\ ,\quad E_{\pm\vec a_{abc}}\ ,\quad E_{\mp\vec
a_7}\ ,\nn\\
&&\vec H'=\vec H -\ft13 (\vec b_7\cdot\vec H)\, \vec b_7 
\ ,\label{d4xx7gen}
\eea
where $i=(a,7)$.  We easily find that these are the generators of
$E_6$, with a regular embedding in $E_7$.  The weights of the $E$
generators under $\vec H'$ are
\be
\pm\vec b_{ab}\ ,\quad \pm\vec a_{abc}\ ,\quad \mp \vec a_7\ .
\ee
The simple roots may be taken to be
\be
\vec\a_1=\vec b_{12}\ ,\quad \vec\a_2=\vec b_{23}\ ,\quad
\vec\a_3=\vec b_{34}\ ,\quad \vec\a_4=\vec b_{45}\ ,\quad
\vec\a_5=\vec b_{56}\ ,\quad \vec\a_6=\vec a_{123}\ .
\ee
The coset parameterising the non-supersymmetric dyonic 2-form
solutions in $D=4$ is therefore given by
\be
\fft{E_7}{E_6}\ ,\label{d42fd}
\ee
which has dimension 55. Note that the $E_6$ denominator group here is
the standard maximally non-compact version $E_{6(6)}$.

     Now we are in a position to discuss the supersymmetry of the
various charge multiplets in $D=4$.  The situation for the 1-charge
orbits whose coset is given by (\ref{d42f1}), is straightforward; these
solutions all preserve $\ft12$ of the supersymmetry.  Similarly, the
2-charge orbits with coset given by (\ref{d42f2}) all preseve $\ft14$
of the supersymmetry, and the 3-charge orbits with coset (\ref{d42f3})
all preserve $\ft18$.  The situation is more complicated for the
4-charge orbits, where the coset is given by (\ref{d42f4}).  In these
cases the solutions either preserve $\ft18$ of the supersymmetry, or
they preserve no supersymmetry at all, depending upon the relative
signs of the four charges.  Of the 16 possible signs, 8 lead to
$\ft18$ supersymmetry, while the other 8 lead to non-supersymmetric
solutions.  (This phenomenon was first seen in the context of the type
IIA string in \cite{lpmult,ko}.  An explanation for why it occurs can
be found in \cite{classp}.)

     A completely different kind of non-supersymmetric solution arises
in the case of the dyonic 2-charge configurations whose orbits are
given by (\ref{d42fd}).  Here, the supersymmetry breaking is quite
independent of the signs of the charges, and the solutions are not
related by any duality symmetries to the non-supersymmetric 4-charge
solutions discussed above.  This can be see from the fact that the
4-charge solutions are described by four harmonic functions, implying
that although non-supersymmetric, there is still no force between the
charges.  By contrast, the 2-charge dyonic solution cannot be
expressed in terms of harmonic functions, and indeed there is a
repulsive force between the two charges \cite{gk}.

\section{Solutions with 1-form field strengths}

     So far, we have discussed the structure of the orbits of
single-charge and multi-charge $p$-brane solutions that are supported
by 4-form, 3-form or 2-form field strengths.  There also exist
$p$-brane solutions that are supported by 1-form field strengths, \ie
by axionic scalar fields.  The situation is much more complicated for
these, because they transform non-linearly under the global symmetry
groups of the fully-dualised versions of the supergravity theories.  
However, different versions of the theories, in which other
dualisation possibilities are implemented, can have different global
symmetries.  For example, if we consider the four-dimensional
supergravity obtained by direct dimensional reduction from $D=11$
without any dualisation, it has an $SL(7,\R)\semi \R^{35}$ global
symmetry, where the $\R^{35}$ corresponds to the shift symmetries of
the 35 axions coming from the dimensional reduction of $A_3$ in
$D=11$.  In this version of $D=4$ supergravity, the 35 axions form a
linear tensor representation under $SL(7,\R)$.  We shall take this an
as example, to illustrate how the stability subgroups depend on the
number of charges in 1-form solutions (strings) supported by these axions. 

     The simple roots for the $SL(7,\R)$ algebra are $\vec b_{i,i+1}$,
for $1\le i\le 6$.  The full set of $SL(7,\R)$ roots are $\pm \vec
b_{ij}$.  As explained above, we shall restrict our attention to the
35 of 1-form field strengths $\F1{ijk}$ that transform linearly under
$SL(7,\R)$.  Their dilaton vectors $\vec a_{ijk}$ form the weights of
the 35 of $SL(7,\R)$.  Let us begin by considering a 1-charge solution
corresponding to the state $\ket{\vec a_{123}}$.  This is annihilated
by the following subset of the $SL(7,\R)$ generators:
\bea
&& E_{\pm\vec b_{12}}\ , \quad  E_{\pm\vec b_{13}}\ , \quad 
 E_{\pm\vec b_{23}}\ , \quad \vec E_{\vec b_{ab}}\ ,\nn\\
&&E_{-\vec b_{1a}}\ ,\quad E_{-\vec b_{2a}}\ ,\quad
E_{-\vec b_{3a}}\ ,\label{d41fgen1}\\
&&\vec H' = \vec H -\ft14 (\vec a_{123}\cdot\vec H)\, \vec a_{123}\ ,
\eea
where we have split the internal index as $i=(1,2,3,a)$.  It is easy
to see that the conjugate pairs of generators, together with the five 
Cartan generators $\vec H'$, give an $SL(3,\R)\times SL(4,\R)$ algebra
(with $E_{\pm\vec b_{ab}}$ associated with $SL(4,\R)$).  The non-zero
roots are
\be
\pm\vec b_{12}\ ,\quad \pm\vec b_{13}\ ,\quad \pm\vec b_{23}\ ,\quad
\pm\vec b_{ab}\ .
\ee
The simple roots may be taken to be
\be
\vec\a_1 = \vec b_{12}\ ,\quad \a_2=\vec b_{23}\ , \quad
\vec \a_3 = \vec b_{45}\ ,\quad \vec\a_4 =\vec b_{56}\ ,\quad
\vec\a_5=\vec b_{67}\ .\label{1root}
\ee
The remaining generators in (\ref{d41fgen1}), form a
mutually-commuting (3,4)-dimensional 
representation under $SL(3,\R)\times SL(4,\R)$.  The highest-weight
generator is $E_{-\vec b_{34}}$, with weight $-\vec b_{34}$ under
$\vec H'$.  Thus the coset describing the 1-charge 1-form solutions is
\be
\fft{SL(7,\R)}{(SL(3,\R)\times SL(4,\R))\semi \R^{12}} \ .
\label{d41f1}
\ee
The dimension of this coset is 13, and all points on the orbit preserve
$\ft12$ of the supersymmetry.

    Turning now to 2-charge solutions, we may consider the case where
these are associated with the states $\ket{\vec a_{123}}$ and
$\ket{\vec a_{145}}$.  We find that the following subset of the
$SL(7,\R)$ generators annihilate this combination of states:
\bea
&&E_{\pm\vec b_{23}}\ ,\quad E_{\pm \vec b_{45}}\ ,\quad E_{\pm\vec
b_{67}}\ ,\quad
(E_{\pm\vec b_{24}}-E_{\mp\vec b_{35}})\ ,\quad (E_{\pm\vec b_{25}}-
E_{\mp\vec b_{34}})\ ,\nn\\
&&E_{-\vec b_{12}}\ ,\quad E_{-\vec b_{13}}\ ,\quad 
E_{-\vec b_{14}}\ ,\quad E_{-\vec b_{15}}\ ,\quad 
E_{-\vec b_{1a}}\ ,\nn\\
&&E_{-\vec b_{2a}}\ ,\quad E_{-\vec b_{3a}}\ ,\quad 
E_{-\vec b_{4a}}\ ,\quad E_{-\vec b_{5a}}\ ,\nn\\  
&&\vec H'=\vec H -\ft14 (\vec a_{123}\cdot\vec H)\, \vec a_{123} -\ft14
(\vec a_{145}\cdot\vec H)\, \vec a_{145}\ ,\label{d41fgen2}
\eea
where we have split the internal index as $i=(1,2,3,4,5,a)$.  The
conjugate pairs of generators, together with three of the four Cartan
generators $\vec H'$, give the
algebra $O(2,3)\times SL(2,\R)$.  The non-zero roots under $\vec H'$ are
\be
\pm\vec b_{23}\ ,\quad \pm \vec b_{45}\ ,\quad \pm\vec b_{67}\ ,\quad
\pm\ft12(\vec b_{24}-\vec b_{35})\ ,\quad \pm\ft12(\vec b_{25}-\vec
b_{34})\ ,
\ee 
and the simple roots may be taken to be
\be
\vec\a_1= \vec b_{23}\ ,\quad \vec \a_2 = \ft12(\vec b_{35}-\vec
b_{24})\ ,\quad \vec \a_3=\vec b_{67}\ .
\ee
($\vec \a_3$ is the simple root for $SL(2,\R)$.)  The remaining Cartan
generator in $\vec H'$, orthogonal to the three simple roots, commutes
with all the generators in $O(2,3)\times SL(2,\R)$, and enlarges the
group to $O(2,3)\times GL(2,\R)$.  The remaining generators in 
(\ref{d41fgen2}) form the
representations (4,1), (1,2) and (4,2) under $O(2,3)\times GL(2,\R)$.
They do not all mutually commute, and the commutator of (4,1) with
(4,2) gives (1,2).  The coset parameterising the 2-charge 1-form
solutions can be written as
\be
\fft{SL(7,\R)}{(O(2,3)\times GL(2,\R))\semi(R^4\oplus R^2 \oplus R^8)}
\label{d41f2}
\ee

   We can in principle proceed to study 1-form solutions with larger
numbers of charges.  In $D=4$, the maximum number of charges for
simple solutions, using the 35 axions $\F1{ijk}$, is 7 \cite{lpsol}.  The
analysis becomes progressively more complicated as the number of
charges increases, and we shall not pursue this further here.
 
\section{Conclusion and discussion}

     In this paper, we have studied the orbits of the single-charge
and multi-charge BPS solutions in maximal supergravities, filled out
by acting with the global symmetry groups.  By identifying the
stability subgroups that leave the original charge configurations
fixed, we obtained the coset spaces that parameterise these orbits.
In some cases, it turns out that a simple single-charge solution
provides an adequate starting point for filling out the entire charge
vector space.  In other cases, such single-charge starting points have
orbits that fail to cover the entire charge vector space.  In these
cases, simple multi-charge solutions exist, which provide more generic
starting points that fill out orbits of larger dimension.   By
studying these orbits in all dimensions $4\le D\le 9$, we obtained the
coset spaces that parameterise the charge vector spaces for all
4-form, 3-form and 2-form solutions.  We also looked at the orbits for
single-charge and 2-charge solutions supported by 1-form field
strengths in $D=4$, under the $SL(7,\R)$ subgroup of $E_7$ under which
the axions coming from the dimensional reduction of $A_3$ in $D=11$
form a linear 35-dimensional representation.

     The procedure that we have followed in this paper is to
investigate how the action of the global symmetry groups of the
maximal supergravities allow us to fill out points in the charge
vector space by starting from basic single-charge or multi-charge
solutions.  As we observed in the introduction, this is different from
the idea of generating multiplets under a spectrum-generating symmetry
that leaves the scalar moduli fixed.  As was shown in a previous paper
\cite{clps}, the standard global supergravity symmetry groups can be
realised on any single-charge family of solutions while holding the
scalar moduli fixed, using a construction in which the moduli are
returned to their initial values by compensating transformations
constructed using the supergravity symmetry's Borel subgroup and the
``trombone'' rescaling transformations, which also act on the metric
and thus rescale the masses.\footnote{Since the maximal compact
subgroup $H$ of the global symmetry group $G$ leaves the scalar moduli
invariant, it might seem more natural to study the orbits of the
multi-charge solutions under $H$.  The problem with doing this is
that, as discussed in \cite{clps}, one cannot make the necessary
discretisation of $H$ that would be needed in order to generate the
discrete orbits of points on the charge lattice that are compatible
with Dirac quantisation.  The subgroup $H$ plus a trombone symmetry
can however be used to generate classical solutions.  The coset
structure of the orbits of multi-charge black hole solutions under the
subgroup $H$ was studied in \cite{ch}.}  Although this construction is
generically non-linear, it nonetheless acts linearly on the charge
vectors of single-charge solutions. For a multi-charge solution,
exactly the same construction can be carried out, taking any one of
the irreducible charge ``components'' of the multi-charge solution as
the basis for the construction. By construction, this will leave the
scalar moduli fixed and the transformation will act linearly on the
chosen irreducible component. Other components of the multi-charge
solution will generally be acted on nonlinearly, in particular as a
result of the trombone rescaling compensator.

     In consequence, the orbits of such fixed-moduli transformations
will not in general cover the full charge space in the same way as we
have discussed in this paper, where the transformations of the scalar
moduli were not taken into account. This point clearly needs more
careful study. Here, we shall be content to give a simple example
taken from $D=9$ supergravity, where the standard supergravity
symmetry is $GL(2,\R)$. One may carry out the construction of
\cite{clps} for the 2-charge solution $(p_1,p_2,q)$ discussed in
section 2, which may be decomposed into irreducible components as
$(p_1,p_2,0)+(0,0,q)$, and selecting, {\it e.g.,} the component
$(p_1,p_2,0)$ as the basis for the construction. On this charge-vector
component, only the $SL(2,\R)$ subgroup of $GL(2,\R)$ is genuinely
active, and on this component the $SL(2,\R)$ transformations act
linearly, by construction of \cite{clps}. On the $(0,0,q)$ component,
however, which would normally be invariant under the standard
$SL(2,\R)$, the trombone rescaling part of the transformation will
cause a rescaling of $q$. In the specific case where the scalar moduli
are all asymptotically vanishing, one will have
$(p_1,p_2,0)\rightarrow (p'_1,p'_2,0)$ according to the standard
$SL(2,\R)$ transformation and $(0,0,q)\rightarrow (0,0,{m'\over m}q)$,
where $m'=\sqrt{{p'}_1^2+{p'}_2^2}$, $m=\sqrt{p_1^2+p_2^2}$. Note that
the orbits of this transformation do not fill out the complete
2-charge space, unlike the discussion in this paper, which ignores the
fixed-modulus constraint.

     In conclusion, we may summarise the distinction between orbits of
spectrum-generating symmetries and the orbits of charges under the
standard supergravity symmetries, ignoring the effect upon the scalar
moduli, as follows.  For the orbits of simple single-charge solutions,
there is no distinction, and so this case is adequately understood.
For simple multi-charge solutions, on the other hand, the
spectrum-generating orbits will generically cover less of the charge
space than those of the standard supergravity symmetries that we have
been discussing in this paper.  The discrepancy occurs because we
currently know only of the single trombone rescaling symmetry that can
serve as a compensator to allow the charges to be moved freely while
holding the scalar moduli fixed.  It is not inconceivable that further
symmetries might exist that would allow one to compensate for changes
in the integration constants occurring in multiple harmonic functions.
If such symmetries were to exist, then the multiplet analysis for
multi-charge solutions that we have given could also be interpreted as
arising from {\it bona fide} spectrum-generating symmetries.  On the
other hand, even if such further symmetries did not exist, the
successful treatment of the orbits for single-charge solutions would
provide as complete a description as is possible for the true
spectrum-generating orbits.  It might in any case be argued to be
somewhat artificial to focus attention on the multi-charge solutions
where the charges are located at the same point in the transverse
space, since the overlapping charges can, at no cost in energy, be
separated.  Thus the true modulus space for multi-charge solutions is
a larger one, in which relative positions and other integration
constants are included also.  Thus one should really be looking for a
symmetry explanation for all the moduli of the multi-charge solutions,
and not only those that determine the asymptotic charges and scalar moduli.

\section*{Note Added} 

      As the work described in this paper was approaching completion,
a paper appeared in which the cosets for BPS solutions in $D=5$ and
$D=4$ were obtained \cite{fg}. The results are substantially in
agreement with ours.  The approach used in \cite{fg} is based on the
Jordan algebras associated with the exceptional groups.

\section*{Acknowledgements}

          We are grateful to Eugene Cremmer and Bernard Julia for helpful 
discussions.  We are grateful for hospitality at SISSA, Trieste.
C.N.P. and K.S.S. are grateful for hospitality at ENS, Paris.

\vfill\eject
\appendix
\section{Cosets for multi-charge $p$-branes}

   In this appendix, we present tables that summarise our results for
the cosets that parameterise the various single-charge and multi-charge
orbits in $4\le D\le 9$.  Table 4 contains the results for $p$-branes
supported by 3-form field strengths, and Table 5 contains the results
for those supported by 2-forms.

\bigskip\bigskip

\centerline{\begin{tabular}{|c|c|c|}\hline
Dimension & $N=1$ & $N=2$ \\ \hline
$D=9$ & $\ffft{SL(2,\sR)}{\sR}$ & -- \\ \hline
$D=8$ & $\ffft{SL(3,\sR)}{SL(2,\sR)\semi \sR^2}$ & -- \\ \hline
$D=7$ & $\ffft{SL(5,\sR)}{SL(4,\sR)\semi \sR^4}$ & -- \\ \hline
$D=6$ & $\ffft{O(5,5)}{O(4,4)\semi \sR^8}$ & $\ffft{O(5,5)}{O(4,5)}$ \\
\hline
\end{tabular}} 
\bigskip\bigskip

\centerline{Table 4: Cosets for $N$-charge 3-form solutions}
\bigskip\bigskip

      The cosets for single-charge 3-form solutions can all be
expressed as $E_{11-D}/(O(10-D, 10-D)\semi \R^p$, where $p=2^{9-D}$.
Thus the denominator group can be viewed as the semi-direct product of
the T-duality group $O(10-D, 10-D)$ with a spinor of $O(10-D, 10-D)$.
This is not surprising since the NS-NS 3-form field strength is a
singlet under T-duality, and is also invariant under the commuting shift
symmetries of all the the R-R axions, which form a spinor
representation under the T-duality group.

\bigskip\bigskip

\centerline{
\begin{tabular}{|c|c|c|c|c|c|}\hline
Dimension & $N=1$ & $N=2$ & $N=3$ & $N=4$ & 
$\begin{array}{c} $N=2$ \\ \hbox{dyon}\end{array}$ \\ \hline
$D=9$ & $\ffft{SL(2,\sR)}{\sR}$ & $\ffft{GL(2,\sR)}{\sR}$ 
                            & -- & -- &--\\ \hline
$D=8$ & $\ffft{SL(3,\sR)\times SL(2,\sR)}{GL(2,\sR)\semi \sR^3}$ &
$\ffft{SL(3,\sR)\times SL(2,\sR)}{SL(2,\sR)\semi \sR^2}$ 
                            & -- & --&--\\ \hline
$D=7$ & $\ffft{SL(5,\sR)}{(SL(3,\sR)\times SL(2,\sR))\semi \sR^6}$ & 
$\ffft{SL(5,\sR)}{O(2,3)\semi \sR^4}$ & -- & --&--\\ \hline
$D=6$ & $\ffft{O(5,5)}{SL(5,\sR)\semi \sR^{10}}$ & 
$\ffft{O(5,5)}{O(3,4)  \semi \sR^8}$& -- & --&-- \\ \hline
$D=5$ & $\ffft{E_6}{O(5,5)\semi \sR^{16} }$ & $\ffft{E_6}{ O(4,5)\semi
\sR^{16}}$ & $\ffft{E_6}{F_4}$ & -- &--\\ \hline
$D=4$ & $\ffft{E_7}{E_6\semi\sR^{27}}$ & $\ffft{E_7}{(O(5,6)\semi
\sR^{32}) \times \sR}$ & $\ffft{E_7}{F_4\semi \sR^{26}}$ &
$\fft{E_7}{E_{6(2)}}$ & $\ffft{E_7}{E_6}$ \\ \hline
\end{tabular}}
\bigskip\bigskip

\centerline{Table 5: Cosets for $N$-charge 2-form solutions}
\bigskip\bigskip

         The denominator group for the single-charge 2-form solution in $D$
dimensions is $E_{10-D}\semi \R^p$, where $p$ is the number of 2-form
field strengths in $D+1$ dimensions.  It is easy to see that this
group will leave the new Kaluza-Klein 2-form invariant.
Since all the 2-form field strengths in $D$ dimensions are equivalent under
$E_{11-D}$, the coset for the Kaluza-Klein 2-form is the same for
any other single-charge solution.   The denominator groups for
for 2-charge solutions also arise with a regular pattern, namely 
$O(9-D,10-D) \semi 
\R^p$, where $p=2^{9-D}$, except for $D=4$, where there is an
additional commuting $\R$ symmetry.  The $\R^p$ is a spinor under 
$O(9-D,10-D)$.

\end{document}